\documentclass[journal]{IEEEtran}
\usepackage{cite}
\usepackage[pdftex]{graphicx}
\usepackage{amsfonts}
\usepackage{amsmath}
\usepackage{array}
\usepackage[caption=false,font=footnotesize]{subfig}
\usepackage{xcolor}
\usepackage{multirow}
\usepackage[english]{babel}

\newcommand{\RM}[1]{\MakeUppercase{\romannumeral #1{}}}

\newcommand{\state}[1]{\textcolor{black}{#1}} 
\newcommand{\causes}[1]{\textcolor{black}{#1}} 
\newcommand{\aim}[1]{\textcolor{black}{#1}} 
\newcommand{\solution}[1]{\textcolor{black}{#1}} 
\newcommand{\outlook}[1]{\textcolor{black}{#1}} 

\begin{document}

\title{Generic Wave Digital Emulation\\ of Memristive Devices}

\author{$^*$Enver Solan and Karlheinz Ochs%
\thanks{E. Solan and K. Ochs are with the Department
of Electrical Engineering and Information Science, Ruhr-University Bochum, Bochum, 44801 Germany e-mail: $^*$enver.solan@rub.de.}%
\thanks{Manuscript received Septemer 14, 2017.}}

\markboth{Journal of \LaTeX\ Class Files,~Vol.~14, No.~8, August~2017}%
{Shell \MakeLowercase{\textit{et al.}}: Bare Demo of IEEEtran.cls for IEEE Journals}

\maketitle

\begin{abstract}
\state{Neuromorphic circuits mimic partial functionalities of brain in a bio-inspired information processing sense in order to achieve similar efficiencies as biological systems. While there are common mathematical models for neurons, which can be realized as nonlinear oscillating circuits, an electrical representation of the synaptic coupling between those is more challenging. Since this coupling strength depends on the learning procedure in the past, it should include a kind of memory. We believe that memristive devices, which are essentially nonlinear resistors with memory, are potential candidates for replacing synapses in neuromorphic circuits.} \causes{Due to a huge number of synapses in a complex neuromorphic circuit, pre-investigations based on simulations of such systems can be very inefficient and time-consuming.} \aim{Flexible and real-time capable memristive emulators, which can directly be incorporated into real circuits, can overcome this problem.} \solution{In our approach, we introduce a generic memristive emulator based on wave digital principles. The proposed emulator is flexible, robust, efficient, and it preserves the passivity of the real device. This, in turn, also makes it reusable independent of a particular application. In the presented work, the emulation of different mathematical models as well as of a real device based on physical models are listed.}
\end{abstract}


\section{Introduction}
 
\state{\IEEEPARstart{I}{t} is widely believed that the efficient working mechanism of the human brain is based on a parallel signal processing approach. Uniformly allocated processing and memory units play a key role in this context. Indeed, the information is stored in the strength of synaptic coupling between neurons, which is also known as synaptic plasticity. In order to achieve similar efficient and powerful signal processing units, neuromorphic circuits, as a part of bio-inspired information processing, try to mimic the brain in terms of an electrical circuit~\cite{indiveri_memory_2015}. While oscillating circuits are the common models for neurons~\cite{behdad_artificial_2015}, an electrical representation of synapses with a learning-dependent coupling strength is more challenging.}

\state{Memristors, which are essentially nonlinear resistors with memory, were postulated from symmetry considerations as the fourth elementary passive circuit device by Chua in 1971~\cite{chua_memristor-missing_1971}. Unique properties of such devices, e.g. the change of its resistance value depending on the current flown through the device in the past, make them to appropriate candidates for modeling synapses in neuromorphic circuits~\cite{ochs_wave_2017,blowers_energy-efficient_2014,ebong_cmos_2012,pershin_neuromorphic_2011,kim_memristor_2012,ochs_wave_2017_02,ochs_anticipation_2017}. Unfortunately, memristors had purely been hypothetical over decades until the first resistive switching device was identified as a memristor in the HP - Hewlett Packard - laboratories~\cite{strukov_missing_2008}. By definition, memristors interrelate electrical charge and magnetic flux. It was generalized to memristive devices and systems in 1976~\cite{chua_memristive_1976}. In contrast to memristors, memristive systems could contain more than one internal state, not being necessarily the flux or charge. Both terms are frequently used in the same context. However, in the scope of this work, we have to distinguish between memristors and memristive systems clearly.}

\causes{Most realized devices can be interpreted as memristive systems. They are fabricated in nanotechnology in order to obtain the memory effect~\cite{hansen_double_2015}. Parameter spread during the fabrication process aggravates reproducible analyses of such devices~\cite{ochs_sensitivity_2016}. Hence, the resulting functionality of a neuromorphic circuit including these devices is also influenced by this parameter spread. Furthermore, the functionality of manufacturable devices is limited by the underlying chemical and physical phenomena within the device. This burdens possible applications, where real devices can be used.}

\state{Mathematical and electrical models for simulation purposes can be utilized to build memristive devices with arbitrary functionalities for reproducible investigations, e.g. in neuromorphic circuits~\cite{biolek_reliable_2013,sah_generic_2015,solan_enhanced_2017}.} \causes{However, biologists suggest that the synaptic density in a human brain is approximately $10^9$ per$~\mathrm{mm}^3$~\cite{nguyen_total_2013}. Even for mimicking a partial functionality of the brain a huge amount of memristive devices is required. This makes the simulations of neuromorphic circuits containing memristive devices inefficient and particularly time-consuming. Besides, simulations are not appropriate for real-time applications, e.g. they cannot be incorporated into real circuits.}

\state{Instead of memristive models in simulations, real-time capable memristive emulators can be utilized to overcome such problems. They can be integrated into a real circuit in order to make reproducible investigations regarding the overall functionality. Real devices can be replaced by such emulators even for a desired functionality which is not manufacturable. Frequently developed memristive emulators are based on hardwired hardware emulators~\cite{hyongsuk_kim_memristor_2012,sanchez-lopez_860khz_2017,biolek_passive_2015}.} \causes{Most often, hardware emulators exhibit a fixed functionality, which means less flexibility. Due to utilized active elements and operational amplifiers, it is often not possible to preserve the passivity of the real device, which is important especially in the context of neuromorphic circuits which belong to self-organizing systems. In this context, ensuring the stability of the overall system is of particular importance. Since the interconnection of subsystems in neuromorphic circuits depends on the learning procedure in the past, stability investigations can be hard because interconnecting stable subsystems lead not necessarily to an overall stable system, whereas the passivity is maintained independently of the interconnection.} \state{Furthermore, it is known that the passivity yields ensured stability properties~\cite{triverio_stability_2007} and, for that reason, passive systems are more convenient in order to achieve stability.} \aim{Regarding these requirements, we intend to introduce a flexible generic memristive emulator while preserving the passivity of the real device, which, in turn, increases the reusability of our approach.} \solution{Therefore, a passive discretization method based on wave digital filters~\cite{fettweis_wave_1986} has been employed. In doing so a robust, efficient and real-time capable algorithmic model of the real device is obtained. It should be emphasized that passivity in this context is understood as an input-output passivity in a digital signal processing sense~\cite{ochs_passive_2001,fettweis_pseudo-passivity_1972}. Due to elementary mathematical operations of the resulting algorithmic model, also called wave digital model, it can be implemented platform-independent as a software emulator on a DSP (digital signal processor), or as a reconfigurable hardware emulator on an FPGA (field programmable gate array) as well as on an ASIC (application specific integrated circuit) considering a hard-wired hardware emulator. Actually, the presented emulation results in this paper are based on a software emulator.}

\begin{figure*}[hbt!]
	\centering
	\includegraphics[scale=1]{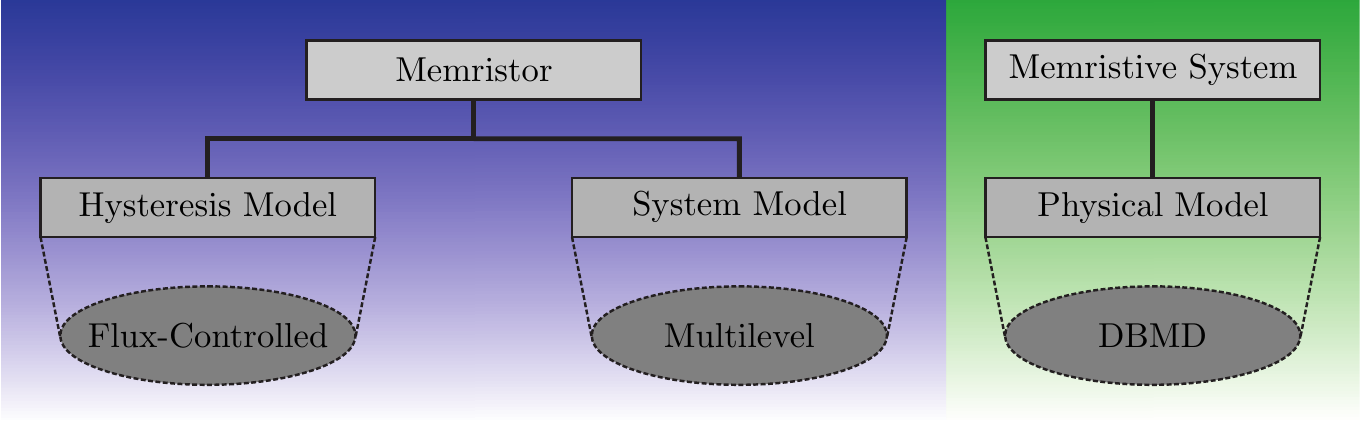}%
	\caption{Categorization of memristive emulators depending on different requirements. Here DBMD - double barrier memristive device - denotes real memristive devices based on interface effects.}%
	\label{fig.:paperStructure}%
\end{figure*}

The paper is organized as illustrated in the block diagram of Fig.~\ref{fig.:paperStructure}: In the next section, a general approach for a wave digital realization of memristive systems is introduced. Afterwards, wave digital emulations of memristors based on a hysteresis as well as system model are shown in Sec.~\ref{sec:Emulation of Memristors}.  Sec.~\ref{sec.:Emulation of Memristive Systems} deals with a physical model in the wave digital framework as an example of a more realistic memristive system.

\hfill mds
 
\hfill September 14, 2017
\\
\section{Wave Digital Emulation of Memristive Systems}
\label{sec.:Wave Digital Emulation of Memristive Systems}%

\subsection{Electrical Representation}
\label{ssec.:Electrical Representation}
\state{In general, memristive devices might possess more than one internal state, not necessarily proportional to the electrical charge or magnetic flux, like for memristors. Regarding the presented devices in this paper, we briefly recapitulate the mathematical description of a voltage-controlled memristive system~\cite{chua_memristive_1976}
\begin{align}
	i(t)
	&= \hat{G}\left(\boldsymbol{z}(t),u(t)\right)\,u(t)\:,\quad
	\dot{\boldsymbol{z}}(t)
	=\boldsymbol{f}\left(\boldsymbol{z}(t),u(t)\right)\:.
	\label{eqn:memristiveSystemElectrical}
\end{align}
The algebraic equation describes the input-output relation with a generalized system response $\hat{G}$, which is the reciprocal value of the memristance, namely memductance. In case of a voltage-controlled device, the description by a memductance instead of a memristance is more appropriate. In the sequel, we use the terminology memristive as well as memductive system depending on the actual mathematical description of the device. In equation~(\ref{eqn:memristiveSystemElectrical}),  the internal state vector is denoted by $\boldsymbol{z}(t)$. The current $i(t)$ is considered as an output, whereas the voltage $u(t)$ is the input signal. The nonlinear vector function $\boldsymbol{f}\left(\boldsymbol{z}(t),u(t)\right)$ is of great importance regarding the memristive behavior of the system. It determines the differential equation with respect to the internal state and has to be integrated in order to get the actual state and with this the memristance (memductance) value over time. Generally, both the memductance $\hat{G}$ as well as the function $\boldsymbol{f}$ can also be time-dependent explicitly.} 

\causes{This general mathematical description can be useful for simulations of general memristive systems~\cite{biolek_reliable_2013}, but it cannot be utilized directly for emulation purposes, at least not for the proposed emulation approach.} \aim{Therefore, an electrical representation of the system as a reference circuit is required.} \state{This task is not trivial and can be challenging for particular memristive systems.}  
\solution{
\begin{figure}[t!]
	\centering
	\subfloat[]{\includegraphics[scale=1]{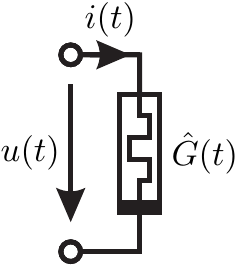}%
		\label{fig.:memristivePortElectrical}}%
	\hfil
	\subfloat[]{\includegraphics[scale=1]{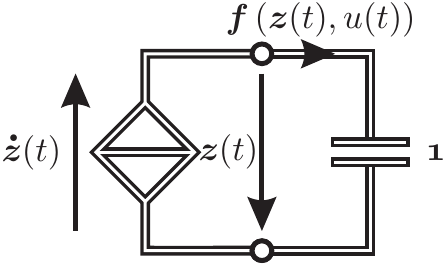}%
		\label{fig.:ElectricalInterpretationMemory}}%
	\caption{Electrical representation of a general memristive system. The input-output relation is interpreted as a memristive one-port (a). The memory of the system is represented by a multidimensional integrator circuit (b).}
	\label{fig.:electrical representation memristive system}
\end{figure}
Fig.~\ref{fig.:electrical representation memristive system} shows such a generalized reference circuit. For the sake of briefness, the state and voltage dependencies of the memristive and memductive systems shown in the following figures are abbreviated by a time-dependency. The electrical representation consists of two parts: the memristive port Fig.~\ref{fig.:memristivePortElectrical} and the electrical interpretation of the memory Fig.~\ref{fig.:ElectricalInterpretationMemory}. The integration of the differential equation describing the memristive behavior is required in order to obtain the internal state $\boldsymbol{z}(t)$. Hence, an integrator circuit represents the electrical interpretation of the memory. Due to the fact that the internal state is, in general, a vector, a multidimensional circuit for this purpose is considered. Regarding a given initial condition $\boldsymbol{z_0}=\boldsymbol{z}(t_0)$ for a starting time $t_0$, the dynamic system described by formula~(\ref{fig.:electrical representation memristive system}) has a unique solution
\begin{align}
	\boldsymbol{z}(t)
	&= \boldsymbol{z_0}+\int_{t_0}^{t}\boldsymbol{f}\left(\boldsymbol{z}(t),u(t)\right) \mathrm{d}\tau\:.
\end{align}}
\state{In case of a scalar memristive function proportional to the input signal, it results in 
\begin{align}
	\dot{z}(t)
	&= \kappa w\left(z(t)\right)u(t)\:,
	\quad \text{with} \quad
	z\in\left(0,1\right)\:,
	\label{eqn.:memristor}
\end{align}
where $w(z)$ denotes typical window functions and $\kappa$ represents a parameter for modeling different material properties. Due to a physically meaningful modeling approach, the normalized state $z$ is restricted between the values $0$ and $1$ by the window function. In order to overcome the boundary lock problem, it should not be equal to $0$ or $1$, because $w(0)=w(1)=0$ and hence $\dot{z}(t)=0$. Applying separation of variables and integrating equation~(\ref{eqn.:memristor}) yields
\begin{align}
	\begin{split}
		z(t)
		&= \lambda^{-1}\left(\kappa\left[\varphi(t)-\varphi_0\right]+\lambda(z_0)\right)\:,\\
		\text{with}\quad
		\lambda(z)
		&= \lambda(z_0)+\int_{z_0}^{z}\frac{1}{w(\tilde{z})}\mathrm{d}\tilde{z}\\
		\quad\text{and}\quad
		\varphi(t)
		&=\varphi_0+\int_{t_0}^{t} u(\tau)\mathrm{d}\tau\:.
	\end{split}
	\label{eqn.:memristor2}
\end{align}
In doing so, the state can be expressed by the magnetic flux and hence the memristor can be interpreted as a special case of a memristive system. Reasonable window functions $w(z)$ are always $>0$. Therefore, the reciprocal function $1/w(z)$ is $>0$ for every $z\in\left(0,1\right)$ and hence $\lambda(z)$ is strictly monotonically increasing so that the inverse function $\lambda^{-1}(z)$ exists for any interval $\left[z_0,z\right]$. It should be mentioned that the magnetic flux in equation~(\ref{eqn.:memristor}) is replaced by the electrical charge for current-controlled devices.}

Since in our approach we are more interested in a memristive emulator instead of a simulation model, the next subsection introduces the method for this purpose.

\subsection{Wave Digital Realization}
\label{ssec:Wave Digital Realization}%
\state{Considering reproducible analyses of complex neuromorphic circuits, memristive emulators are much more appropriate than simulation models. Significant requirements have to be taken into account to incorporate such emulators into real circuits.} \aim{Considering the reusability of emulators, independent of a particular application, energetic properties of the real device, like passivity, has to be maintained.} \state{Especially for neuromorphic circuits, where the interconnection of several subsystems depends on the learning process in the past, this point is of particular importance. Most presented memristive emulators are based on hardware emulators including operational amplifiers and other active elements like current conveyors~\cite{sanchez-lopez_860khz_2017,hyongsuk_kim_memristor_2012}. Hence they can be built by common electrical components.} \causes{Unfortunately, such emulators match not energetic properties of the real physical device, at least not the passivity criterion. It is known that the interconnection of stable sub-systems leads not necessarily to a stable overall system. In contrast to that, the interconnection of passive sub-systems always yields a passive overall system. Furthermore, stability conditions are satisfied by passive systems~\cite{triverio_stability_2007}.} \aim{Consequently, passive systems yield convenient stability investigations in a system theoretic sense, although passivity means not necessary stability. This means that the passivity property formulates a significant requirement regarding reusable emulators.} \state{Indeed,~\cite{biolek_passive_2015} presents a passive memristive emulator.} \causes{Although the emulator is passive, it contains a limited programmability of the memristive function.} \aim{In order to emulate memristive systems considering different applications, it should also be flexible even during runtime.} 

\solution{Here, we introduce a flexible emulator of memristive systems by utilizing a passive discretization method based on wave digital principles. The wave digital method was originally introduced to get a digital replica of analog filters~\cite{fettweis_wave_1986}. Several benefits of this approach make it suitable for digitizing memristive systems. For example, the resulting algorithmic model preserves the passivity of the real physical device~\cite{fettweis_pseudo-passivity_1972}. Passivity in this context should be understood as an input-output passivity in a digital signal processing sense. This clearly means that the passivity in this context is based on the discrete definition of the power supplied to or drawn from the system so that the discrete numerical values processed by the algorithmic model can hold a corresponding physical meaning.
\begin{figure}[t!]
	\centering
	\includegraphics[scale=1]{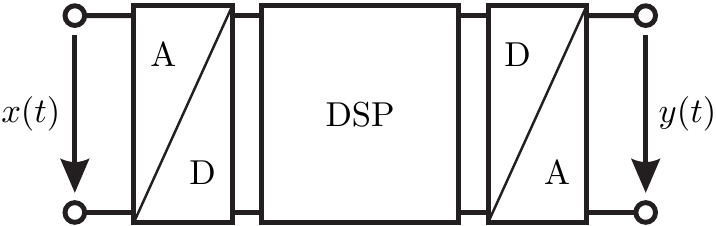}%
	\caption{Illustrative sketch for the implementation of an algorithmic model of a real physical system on a digital signal processor (DSP) based on wave digital principles, with input $x(t)$ and output $y(t)$.}%
	\label{fig.:implementationSketch}%
\end{figure}
Furthermore, the wave digital algorithmic model consists of elementary mathematical operations like adders, multipliers and delay elements. This leads to a platform-independent implementation, e.g. as a software emulator on a digital signal processor (DSP), cf. Fig.~\ref{fig.:implementationSketch}. In doing so, a flexible emulator is achieved by programming a desired memristive behavior. Moreover, the wave digital method yields robust, efficient and real-time capable algorithms, which can additionally be exploited for investigations on a simulation level or for real-time digital signal processing approaches. A detailed recapitulation of the wave digital theory goes far beyond the scope of this paper, and the interested reader must be referred to~\cite{fettweis_wave_1986}.}

\state{The wave digital method basically includes two steps in order to get an algorithmic model representing the physical device. Firstly, the time-discretization transfers the underlying differential equations into difference equations. Secondly, replacing voltages and currents by wave quantities, known from the scattering theory, as signal parameters leads to explicit equations, which are more suitable for implementations. The relationship between voltages, currents, and wave quantities is given by a bijective mapping
\begin{align}
	a
	=u+R\,i
	\quad \text{and} \quad
	b
	=u-R\,i\:,
	\quad\text{with}\quad
	R>0\:.
	\label{eqn.:bijectiveMapping}
\end{align}
Here, $a$ and $b$ denote the incident and reflected wave at a particular port, respectively. The corresponding port resistance $R$ is an arbitrary positive constant.} 

\solution{In the following, electrical representations of memristive systems are used as reference circuits in order to apply the wave digital method. Regarding memristive systems described by~(\ref{eqn:memristiveSystemElectrical}), the time discretization $t_k=t_0+kT$, $k\in\mathbb{N}$, with sampling period $T>0$, leads to 
\begin{align}
	i(t_k)
	&= \hat{G}\left(\boldsymbol{z}(t_k),u(t_k)\right)u(t_k)
\end{align}
for the memristive one-port described by the algebraic equation. Replacing voltages and current by wave quantities yields
\begin{align}
	\begin{split}
		b(t_k)
		&= \rho\left(\hat{G}\left(\boldsymbol{z}(t_k),\frac{a(t_k)+b(t_k)}{2}\right)\right)\,a(t_k)\:,\\
		&\text{with}\quad
		\rho(\hat{G})
		= \frac{1-R\hat{G}}{1+R\hat{G}}\:,
	\end{split}
\label{eqn.:memristiveSystemWd}
\end{align}
where $\rho$ denotes a reflection coefficient from scattering parameter theory.
\begin{figure}[!t]
	\centering
	\subfloat[]{\includegraphics[scale=0.7]{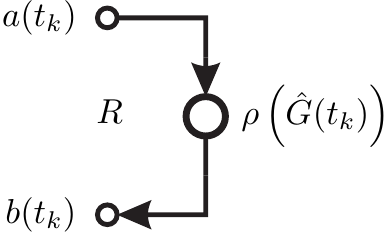}%
		\label{fig.:memristor_wd}}
	\hfil
	\subfloat[]{\includegraphics[scale=0.7]{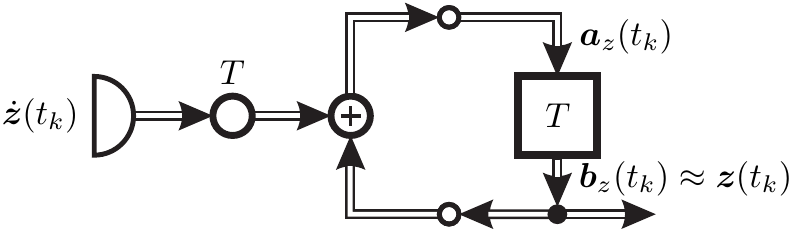}%
		\label{fig.:integrator_wd}}
	\caption{Wave digital representation of a memristive system with memristive reflection coefficient (a) and corresponding wave digital realization of the memory (b).}
	\label{fig.:memristivesystem_wd}
\end{figure}
Due to a variable memristance (memductance), the reflection coefficient also varies over time, cf. Fig.~\ref{fig.:memristor_wd}.}

\solution{The remaining part is the electrical interpretation of the memory. For this, the integrator circuit, which is shown in Fig.~\ref{fig.:ElectricalInterpretationMemory}, is utilized as a reference circuit. Starting from an instance $t_{k-1}$ we get at instance $t_k$ the actual state sample
\begin{align}
	\boldsymbol{z}(t_k)
	&= \boldsymbol{z}(t_{k-1})+\int_{t_{k-1}}^{t_k}\boldsymbol{f}\left(\boldsymbol{z}(\tau),u(\tau)\right)\mathrm{d}\tau\:.
\end{align}
For a wave digital realization, this integration has to be done numerically, and it is common to use the trapezoidal rule
\begin{align}
	\begin{split}
		\boldsymbol{z}(t_k)
		&\approx 			\boldsymbol{z}(t_{k-1})+\frac{T}{2}\left[\boldsymbol{\dot{z}}(t_k)+\boldsymbol{\dot{z}}(t_{k-1})\right]\:,\\
		&\text{with}\quad
		\boldsymbol{\dot{z}}(t_k)
		=\boldsymbol{f}\left(\boldsymbol{z}(t_k),u(t_k)\right)\:.
	\end{split}
\label{eqn.:numericalIntegration}
\end{align}
The reinterpretation of the state as the resulting voltage over the capacitance, the time derivative of the state, namely the memristive function $\boldsymbol{f}$, as the current through it in Fig.~\ref{fig.:ElectricalInterpretationMemory}, and applying the bijective mapping~(\ref{eqn.:bijectiveMapping}) lead to
\begin{align}
	\begin{split}
		\boldsymbol{b}_z(t_k)
		&= \boldsymbol{z}(t_k)-\frac{T}{2}\boldsymbol{\dot{z}}(t_k)\\
		&\approx \boldsymbol{z}(t_{k-1})+\frac{T}{2}\boldsymbol{\dot{z}}(t_{k-1})
		= \boldsymbol{a}_z(t_{k-1})\:,
	\end{split}
\end{align}
where $R=T/2$ is assumed as the port resistance. From the limit 
\begin{align}
	\lim_{T\rightarrow 0}\boldsymbol{z}(t_k)=\boldsymbol{b}_z(t_k)
\end{align}
it follows that for small step sizes $\boldsymbol{z}(t_k)\approx\boldsymbol{b}_z(t_k)$ is a good approximation and leads to the wave digital realization of Fig.~\ref{fig.:integrator_wd}, considering the wave digital realization of an ideal current source
\begin{align}
	\boldsymbol{a}_z(t_k)
	&= \boldsymbol{f}(\boldsymbol{z}\left(t_k\right),u\left(t_k\right))T+\boldsymbol{b}_z(t_k)\:.
\end{align}
With this approximation, the storage in Fig.~\ref{fig.:integrator_wd} can be initialized approximatively by the initial state $z_0$.}

\causes{At first glance, one may overlook that equation~(\ref{eqn.:memristiveSystemWd}) is implicit with respect to $b$. As a consequence, the signal flow graph in Fig.~\ref{fig.:memristor_wd} contains an invisible algebraic loop.} 
\solution{
\begin{figure*}[t!]
	\centering
	\subfloat[]{\includegraphics[scale=0.9]{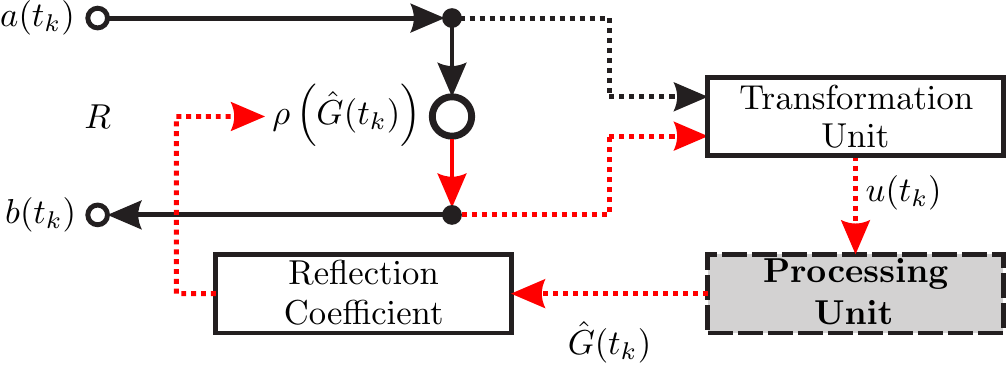}%
		\label{fig.:extended wave flow diagram}}%
	\hfil
	\subfloat[]{\includegraphics[scale=0.9]{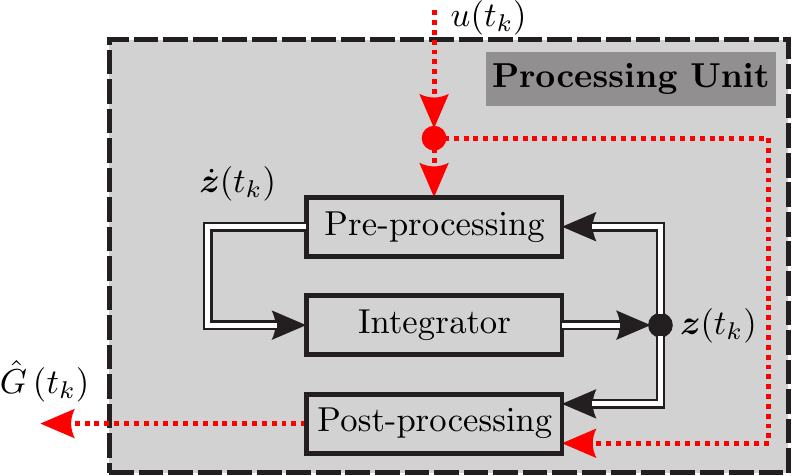}%
		\label{fig.:processing_unit}}%
	\caption{Wave digital realization of a memristive device by a variable reflection coefficient calculated through an extended wave flow diagram (a) and processing unit (b). The invisible algebraic loop is depicted as the red dashed curve.}
	\label{fig.:waveDigitalRealizationMemristiveDevice}
\end{figure*}
Fig.~\ref{fig.:extended wave flow diagram} illustrates the missing algebraic loop. There, the dashed lines indicate the extension, which contains three functional blocks: A transformation unit, a processing unit, and the computation unit of the reflection coefficient. The red-colored delay-free directed loop visualizes the aforementioned implicit part of equation~(\ref{eqn.:memristiveSystemWd}). For a digital signal processing, it is important that this implicit equation has to be solved at each instance. To this end, one can use a fix-point iteration, which requires repeating the computations of the red path several times while the signals on the black paths remain constant. It should be stressed that incorporating an iterative process into the wave digital model destroys not the overall passivity of the resulting algorithmic model~\cite{schwerdtfeger_multidimensional_2014}. In case of a high sampling rate, the signals vary slowly from instance to instance and the number of iterations can be significantly reduced.}

\outlook{Apart from this argument of sufficient numerical accuracy, there is an additional physical point of view. For this, we focus on a memristor with ion drift. If the input voltage changes, the ions of the memristor cannot move instantly to the new position because of their mass moment of inertia. In such a case, the implicit equation is a kind of modeling error and not to iterate is a physically more accurate approach. In this way, the signal-processing algorithm is suitable for real-time applications.}

\solution{As shown in Fig.~\ref{fig.:extended wave flow diagram}, within every iteration step, the signals are processed by three units. The transformation unit maps the waves $a$ and $b$ onto the voltage $u$ subject to (\ref{eqn.:bijectiveMapping}). The processing unit determines from this voltage the state-dependent memductance $\hat{G}$, which is used by a computation unit to compute the reflection coefficient compliant with (\ref{eqn.:memristiveSystemWd}). This unit is subdivided into further three units as depicted in Fig.~\ref{fig.:processing_unit}. Here, the pre-processing unit contains the memristive function $\boldsymbol{f}$. In order to get the actual state, this function has to be integrated, which is done by the integrator, cf. Fig~\ref{fig.:ElectricalInterpretationMemory} and Fig.~\ref{fig.:integrator_wd}. Finally, the post-processing unit evaluates the memductance with respect to the actual state.}

In the sequel, the presented emulators are directly coupled to a resistive voltage source, with internal resistance $R_0=0.1\Omega$. 
\begin{figure}[!t]
	\centering
	\subfloat[]{\includegraphics[scale=0.7]{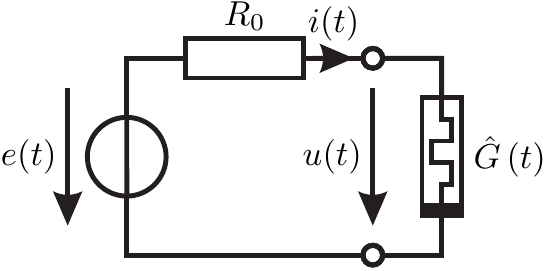}%
		\label{fig.:electricalRepresentationScenario}}
	\hfil
	\subfloat[]{\includegraphics[scale=0.8]{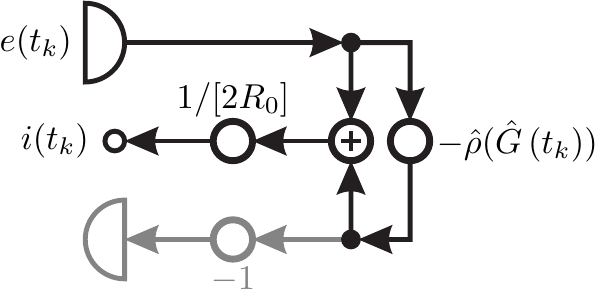}%
		\label{fig.:waveDigitalRepresentationScenario}}
	\caption{Electrical circuit for emulation validation (a) and corresponding wave flow diagram (b).}
	\label{fig.:emulationScenario}
\end{figure}
The electrical circuit is shown in Fig.~\ref{fig.:electricalRepresentationScenario} and the corresponding wave digital realization is depicted in Fig.~\ref{fig.:waveDigitalRepresentationScenario}. Assuming the current as output, the gray path of Fig.~\ref{fig.:waveDigitalRepresentationScenario} is not needed here. It should be emphasized that, in the presented work, results have been obtained from a software emulator.

\section{Emulation of Memristors}
\label{sec:Emulation of Memristors}%
\aim{This section aims to present two possibilities for wave digital memristor emulations based on different requirements. For the first approach, a hysteresis model is given. This means, that for an arbitrary given functionality, e.g. in form of a hysteresis curve, an emulator, which matches exactly this functionality is desired. In the second example, the overall functionality of a more complex system including memristors is given. Considering the functionality of the overall system, a memristor emulator for achieving such a functionality should be modeled and emulated. We will show that both application examples are covered by the generic wave digital method.}

\subsection{Hysteresis Model}
\label{ssec.:Hysteresis Model}%
\state{The hysteresis model introduces a method in order to emulate memristors with a desired functionality given by their hysteresis curves.} \causes{Since the hysteresis changes for different input signals, it cannot be considered as a unique characterization of a memristor.} \aim{In order to get an emulator, which mimics exactly the same hysteresis curve, an identification procedure based on a measured or simulated hysteresis is required~\cite{ochs_wave_2016_02}.} 

\state{A memductor (memristor) is uniquely characterized by its memductance over flux map $\hat{G}(\varphi)$ (memristance over charge map $\hat{R}(q)$), cf.~\cite{chua_resistance_2011}.} \solution{Once this curve is obtained, it can be implemented in the post-processing unit of Fig.~\ref{fig.:processing_unit} in order to get a wave digital emulator, which mimics the required hysteresis curve and responds also with respect to other input signals.}

\solution{
\begin{figure}[!t]
	\centering
	\includegraphics[scale=0.66]{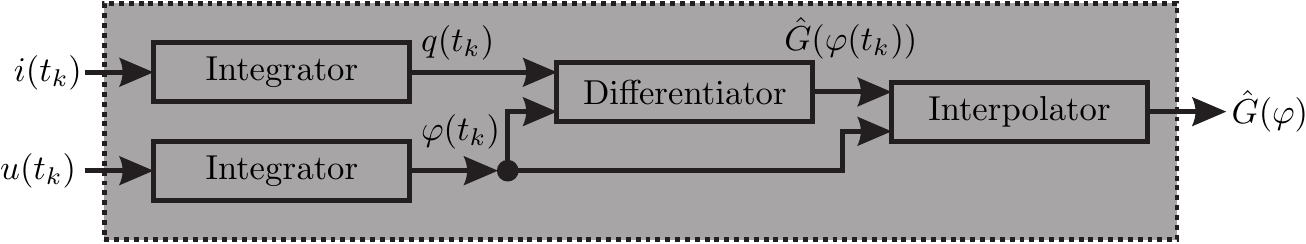}
	\caption{Block diagram of the identification procedure.}
	\label{fig.:identification}%
\end{figure}
The block diagram in Fig.~\ref{fig.:identification} shows a possibility to derive a memductance function $\hat{G}(\varphi)$ in case of uniform sampled values from current~$i(t_k)$ and voltage~$u(t_k)$ of a memristor. Numerically integrating the voltage and current leads to samples for the charge~$q(t_k)$ and flux~$\varphi(t_k)$ , respectively. The latter samples in turn are used to achieve an approximation of the memductance samples~$\hat{G}(\varphi(t_k))$ by exploiting the constitutive relation~\cite{chua_resistance_2011}
\begin{align}
	\hat{G}(\varphi)
	&= \frac{\mathrm{d}\hat{q}(\varphi)}{\mathrm{d}\varphi}
\end{align}
and an appropriate numerical differentiation, e.g. forward difference quotient. In this manner, the associated samples of flux~$\varphi(t_k)$ and memductance values~$\hat{G}(\varphi(t_k))$ can suitably be interpolated in order to obtain the desired flux-dependent memductance function~$\hat{G}(\varphi)$. Although a linear interpolation has been exploited in this work, every kind of other interpolation techniques can be utilized in this context.}

\outlook{This idea can be extended to non-uniform sampled values of current and voltage by interpolating those sampling values at first and uniformly sampling them afterwards. Moreover, it is of no matter where the samples originate from. For instance, we can use measured values of a physical real memristor as well as simulation results considering memristor models with distributed parameters, or a desired memristor behavior. In order to increase the efficiency of a real-time implementation, the memductance of real or simulated memristors can also be approximated by curve fitting techniques. To conclude, all kinds of memristors, independent of the initial state, can be emulated.} 

In the following, some examples of the identification procedure considering varying kinds of memristors are demonstrated. Therefore, an arbitrary hysteresis curve for a sinusoidal input signal 
\begin{align}
	e_\mathrm{s}(t)
	&= E\sin\left(2\pi Ft\right)\:
	\label{eqn.:sinusoidalInput}
\end{align}
is given with voltage amplitude $E$ and frequency $F$. From this, we derive the flux
\begin{align}
	\varphi(t) 
	&= \Phi_\mathrm{s} \sin^2(\pi F t)\:,
	\ \text{with} \
	\varphi(0) = 0\:,\:
	\Phi_\mathrm{s} = \frac{E}{\pi F}\:.
	\label{eqn:scenario flux}%
\end{align}
In order to identify the memristor, we have to bear the dynamic range of the identification in mind. Because of the utilized excitation, the flux has a limited amplitude, which also restricts the range of validity regarding the identification to~$\varphi\in[0,\Phi_s]$. Considering this limitation, a periodic triangular voltage
\begin{align}
	e_\Delta(t)
	&= \frac{2E}{\pi}\arcsin\left(\sin\left(2\pi Ft\right)\right)\:
	\label{eqn.:triangular input}
\end{align}
is used to excite the virtualized memristor with the same amplitude and period as in~(\ref{eqn.:sinusoidalInput}). Then, one has $\Phi_v/\Phi_s=\pi/4<1$, where $\Phi_v$ denotes the amplitude of resulting flux for input voltage $e_\Delta(t)$. It is guaranteed that the flux of this excitation voltage remains inside the range of validity. 

\subsubsection{Binary Switch Memristor}
\state{Memristors with a binary switching property change their resistance value between two discrete values and hence they are appropriate devices in nonvolatile memory applications~\cite{ho_dynamical_2011} regarding a storage of one-bit information. Physically, such devices are often realized as electrochemical metalization cells (ECM-cells)~\cite{dirkmann_kinetic_2015}.} \causes{It is hard to develop devices in nanotechnology that match the requirements exactly for a nonvolatile memory~\cite{waser_redox-based_2009}.} \aim{An emulator might be used to make previous investigations before fabrication.} \solution{
	\begin{figure*}[t!]
		\centering
		\includegraphics[scale=1]{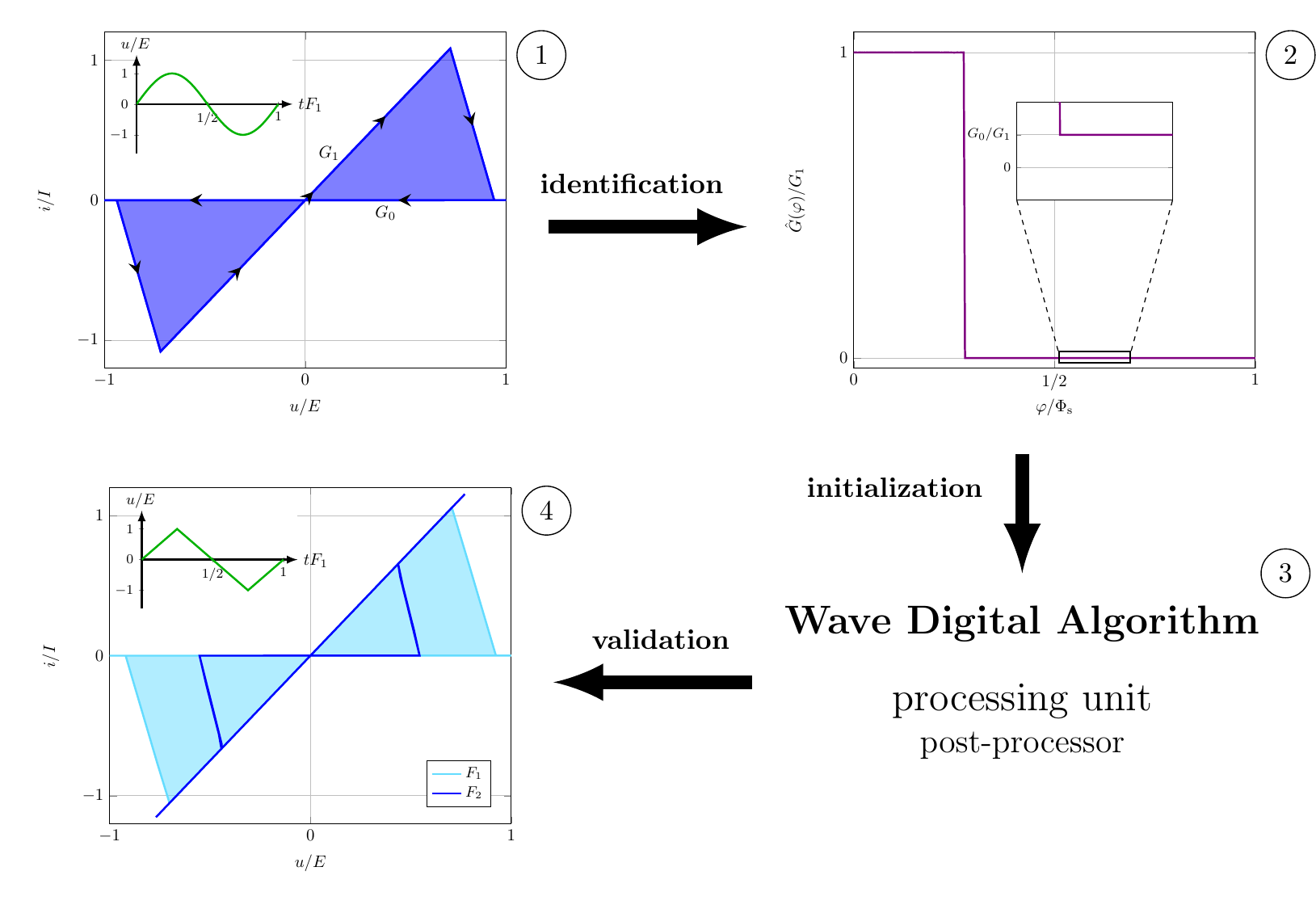}
		\caption{Illustrated workflow for a wave digital emulation of an identified memristor. From the desired functionality $1$, corresponding to a particular input signal, the characteristic curve $2$ is determined and implemented into the wave digital framework $3$ in order to emulate the memristor for arbitrary input signal $4$.}%
		\label{fig.:identificationBinarySwitch}%
	\end{figure*}
The whole procedure in order to get an emulator of a binary switching memristor is illustrated in Fig.~\ref{fig.:identificationBinarySwitch}. There, a typical (given) hysteresis curve of such a device is depicted in Fig.~\ref{fig.:identificationBinarySwitch}-$1$. As it can be seen, the memductance changes between the high $G_1$ and low conductance state $G_0$, respectively. Applying the identification procedure as shown in Fig.~\ref{fig.:identification} leads to the characteristic curve of the device, cf. Fig~\ref{fig.:identificationBinarySwitch}-$2$. In order to emulate the memristor with the wave digital approach, the post-processing unit in Fig.~\ref{fig.:processing_unit} of the wave digital framework has been initialized by this characteristic curve. The emulator is validated by exciting it with two triangular input signal with different periods. As it can be seen from Fig.~\ref{fig.:identificationBinarySwitch}-$4$, the hysteresis area decreases for higher frequencies, which is a typical fingerprint of a memristor.} Utilized parameters in order to get the presented results are centralized in Tab.~\RM{1}.
\begin{table}[h!]
	\centering
	\renewcommand{\arraystretch}{1.3}
	\caption{Parameters binary switch memristor}
	\label{tab:one}
	\begin{tabular}{l r@{\,} r@{\,} l@{\,} l@{\,}}
		\hline
		\multicolumn{5}{c}{Software Emulation} \\
		Starting time        & $t_0$ & $=$ & $0$ & $\mathrm{s}$ \\
		Sampling period      & $T$   & $=$ & $1$ & $\mathrm{ms}$\\
		Number of iterations & $n_i$ & $=$ & $1$ &              \\
		\hline
		\hline
		\multicolumn{5}{c}{Excitation signals} \\
		Amplitude   & $E$   & $=$ & $5$ & $\mathrm{V}$ \\
		Frequency 1 & $F_1$ & $=$ & $1$ & $\mathrm{Hz}$\\
		Frequency 2 & $F_2$ & $=$ & $2$ & $\mathrm{Hz}$\\
		\hline
		\hline
		\multicolumn{5}{c}{Memristor} \\
		High memductance state & $G_1$ & $=$ & $3$   & $\mathrm{S}$   \\
		Low memductance state  & $G_0$ & $=$ & $100$ & $\mu\mathrm{S}$\\
		Normalization current  & $I$   & $=$ & $10$  & $\mathrm{A}$   \\
		\hline
	\end{tabular}
\end{table}

\subsubsection{Memristor with a continuous resistance range}
\state{Memristors or memristive devices are regarded as potential candidates for replacing synapses in neuromorphic circuits~\cite{thomas_memristor-based_2013}.} \causes{Due to the fact that the synaptic weighting is of continuous nature, binary switching devices are less appropriate for such applications.} \aim{For utilizations in neuromorphic circuits, devices with a continuous resistance range are suitable.} \solution{In contrast to the previous example, in this case, a memristor with a continuous resistance range is exploited in the identification approach.
\begin{figure}[t!]
	\centering
	\includegraphics[scale=0.8]{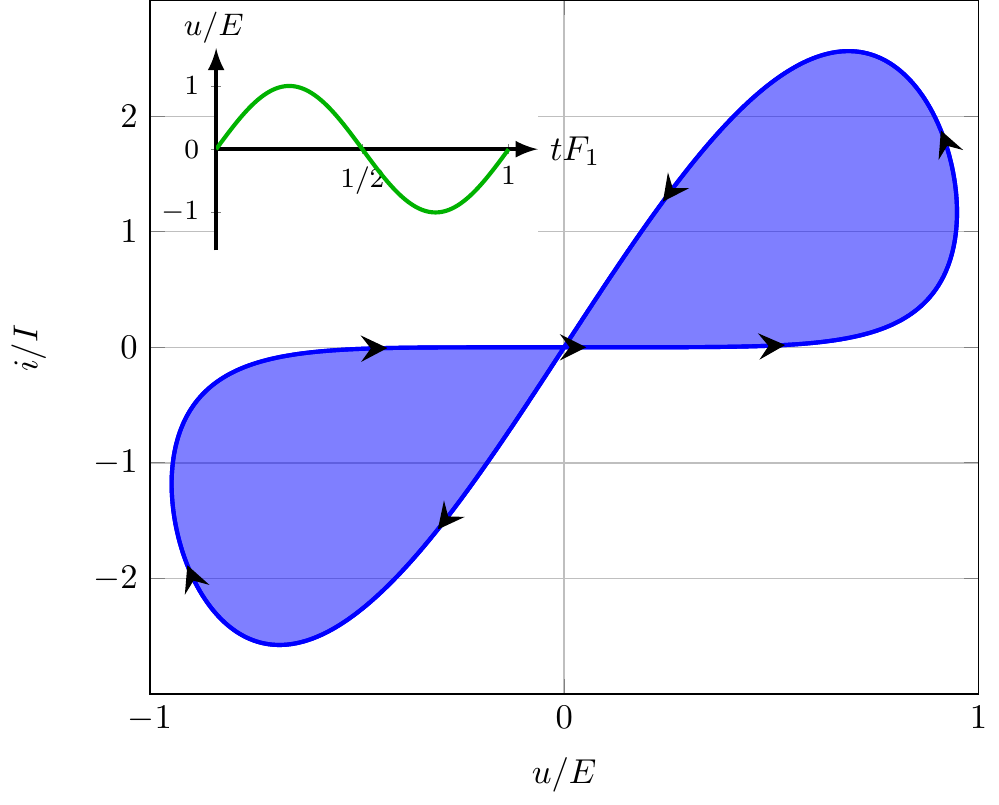}%
	\caption{Typical hysteresis curve of a memristor with a continuous resistance range.}%
	\label{fig.:continuousHysteresisModel}%
\end{figure}
\begin{figure}[t!]
	\centering
	\includegraphics[scale=0.8]{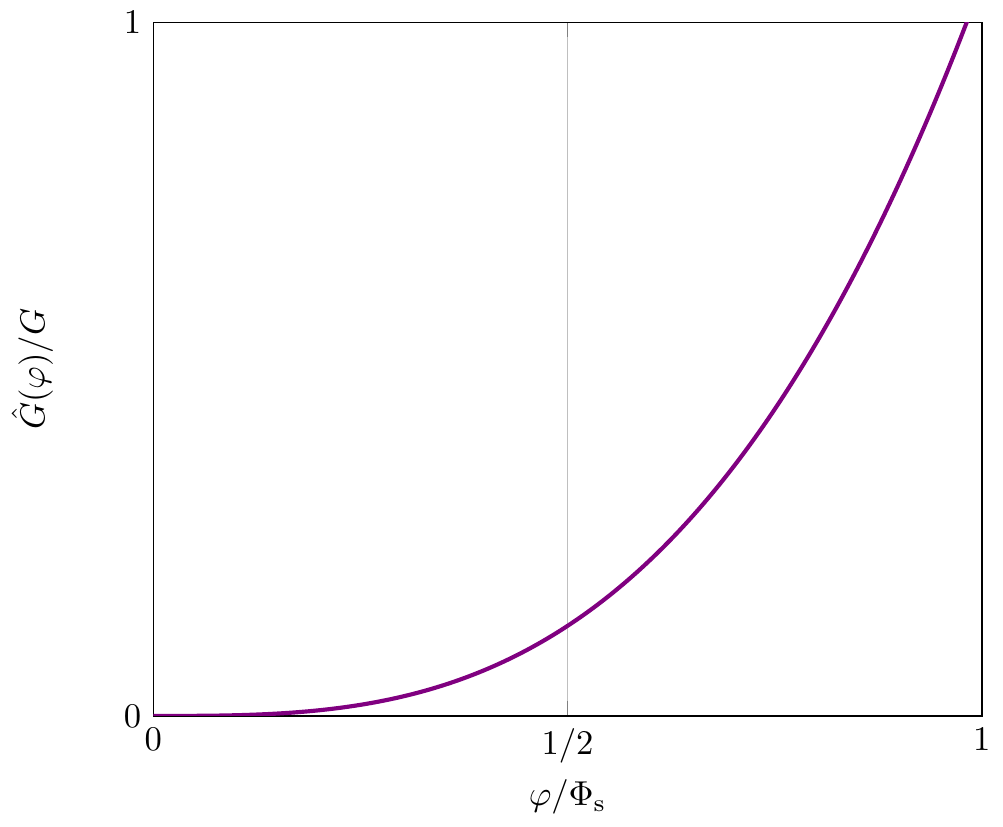}%
	\caption{Characteristic curve of the device with a continuous resistance range.}%
	\label{fig.:continuousIdentification}%
\end{figure}
A typical hysteresis curve of such a device is shown in Fig.~\ref{fig.:continuousHysteresisModel}, which leads to the characteristic curve depicted in Fig.~\ref{fig.:continuousIdentification}.
\begin{figure}[t!]
	\centering
	\includegraphics[scale=0.8]{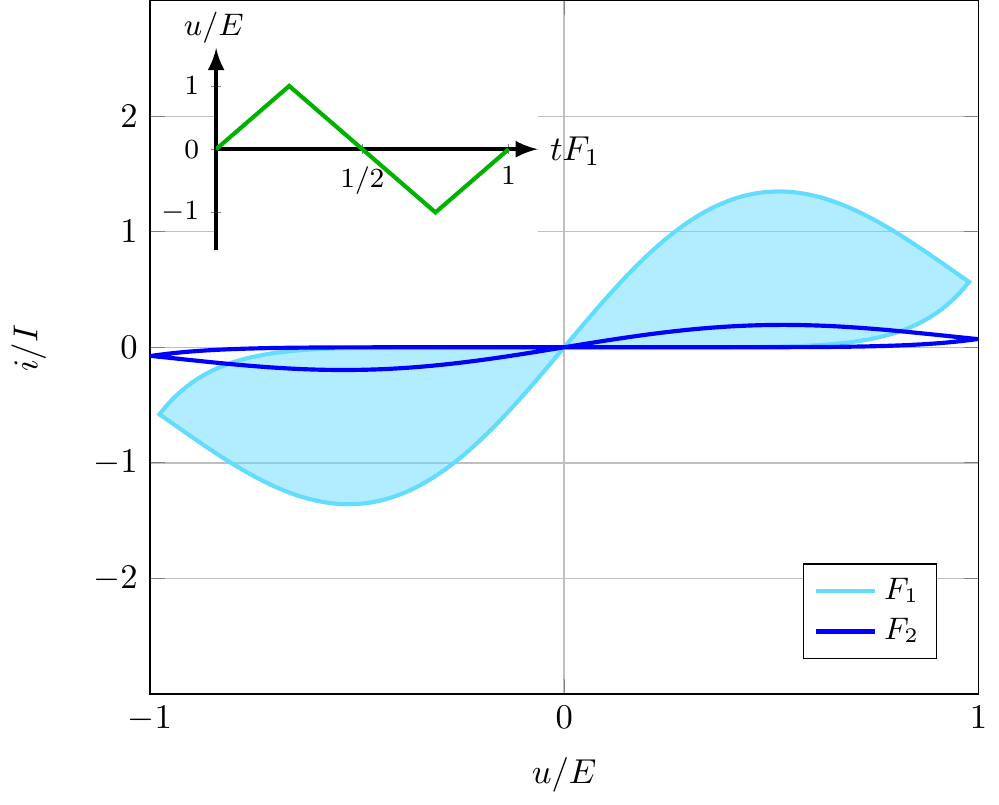}%
	\caption{Wave digital emulation of the identified memristor with a continuous resistance range for a triangular input signal with two different periods.}%
	\label{fig.:continuousValidation}%
\end{figure}
Similar to the illustration in Fig.~\ref{fig.:identificationBinarySwitch}, a wave digital implementation of this characteristic curve into the wave digital framework yields the hysteresis curves of Fig.~\ref{fig.:continuousValidation}, considering two frequencies $F_1$ and $F_2$. As excepted, the dependency of the hysteresis area with respect to the excitation frequency can be obtained.} The utilized parameters and their corresponding values are listed in Tab.~\RM{2}.
\begin{table}[h!]
	\centering
	\renewcommand{\arraystretch}{1.3}
	\caption{Parameters continuous memristor}
	\begin{tabular}{l r@{\,} r@{\,} l@{\,} l@{\,}}
		\hline
		\multicolumn{5}{c}{Software Emulation} \\
		Starting time        & $t_0$ & $=$ & $0$ & $\mathrm{s}$ \\
		Sampling period      & $T$   & $=$ & $1$ & $\mathrm{ms}$\\
		Number of iterations & $n_i$ & $=$ & $1$ &              \\
		\hline
		\hline
		\multicolumn{5}{c}{Excitation signals} \\
		Amplitude   & $E$   & $=$ & $5$ & $\mathrm{V}$ \\
		Frequency 1 & $F_1$ & $=$ & $1$ & $\mathrm{Hz}$\\
		Frequency 2 & $F_2$ & $=$ & $2$ & $\mathrm{Hz}$\\
		\hline
		\hline
		\multicolumn{5}{c}{Memristor} \\
		Normalization conductance  & $G$   & $=$ & $3$  & $\mathrm{S}$   \\
		Normalization current  & $I$   & $=$ & $2$  & $\mathrm{A}$   \\
		\hline
	\end{tabular}
\end{table}

\subsubsection{HP ion drift model}
\state{There are many mathematical models of memristors in the literature. A prominent one is the HP-ion drift model with nonlinear dopant drift~\cite{biolek_reliable_2013}
\begin{align}
	\begin{split}
		&u(t)
		= \hat{R}\left(z(t)\right)i(t)\:,\:
		\hat{R}(z)
		= R_0z+R_1\left[1-z\right]\:,\\
		&\text{with}\:
		\dot{z}(t)
		= \kappa w\left(z(t)\right)i(t)
		\:\text{and}\:
		w(z)
		= 1-\left[2z-1\right]^{2p}\:,
	\end{split}
	\label{eqn.:hpIonDriftModel}
\end{align}
where $w(z)$ is a window function and restricts the internal state in the domain $z\in\left[0,1\right]$, cf. equation~(\ref{eqn.:memristor}). The material properties are incorporated in $\kappa$, whereas $R_0$ and $R_1$ are the low and high resistance states, respectively. With the parameter $p$, the steepness of the window function at the boundaries can be controlled. It can be observed in equation~(\ref{eqn.:hpIonDriftModel}) that the time derivative of the state is proportional to the input signal and, consequently, the model describes a memristor, cf. equation~(\ref{eqn.:memristor}), which can, in turn, be identified and emulated by our wave digital approach. Here, the input signal is the current, whereby equation~(\ref{eqn.:hpIonDriftModel}) represents a current-controlled memristor. By applying partial fraction decomposition before integration, we identify for $p=1$ the function
\begin{align}
	\lambda_{1}(z)
	&= \frac{1}{4\kappa}\ln\left(\frac{z}{1-z}\right)\:,
	\quad\text{assuming}\quad
	z\in\left(0,1\right)\:.
	\label{eqn.:lambdaHP}
\end{align}
With subject to equation~(\ref{eqn.:memristor2}), the state can be formulated in dependence of the electrical charge. Incorporating the resulting expression into the original memristor model with $q_0=0$ leads to, cf.~\cite{biolek_reliable_2013}:}
\begin{align}
	\begin{split}
		&\hat{R}(z)
		= \hat{R}\left(\lambda_{1}^{-1}\left(q(t)+\lambda_{1}^{-1}(z_0)\right)\right)\\
		&= R_1+\frac{R_0-R_1}{1+\gamma e^{-4\kappa q(t)}}\:,
		\quad\text{with}\quad
		\gamma
		= \frac{\hat{R}(z_0)-R_0}{R_1-\hat{R}(z_0)}\:.
	\end{split}
	\label{eqn.:hpModelChargeDependent}
\end{align}
\aim{In order to emulate such an memristor, we can directly initialize the post-processing unit with the mathematical model~(\ref{eqn.:hpModelChargeDependent}). However, we are more interested in the identification of the device by measurements or simulations. 
\begin{figure}[t!]
	\centering
	\includegraphics[scale=0.8]{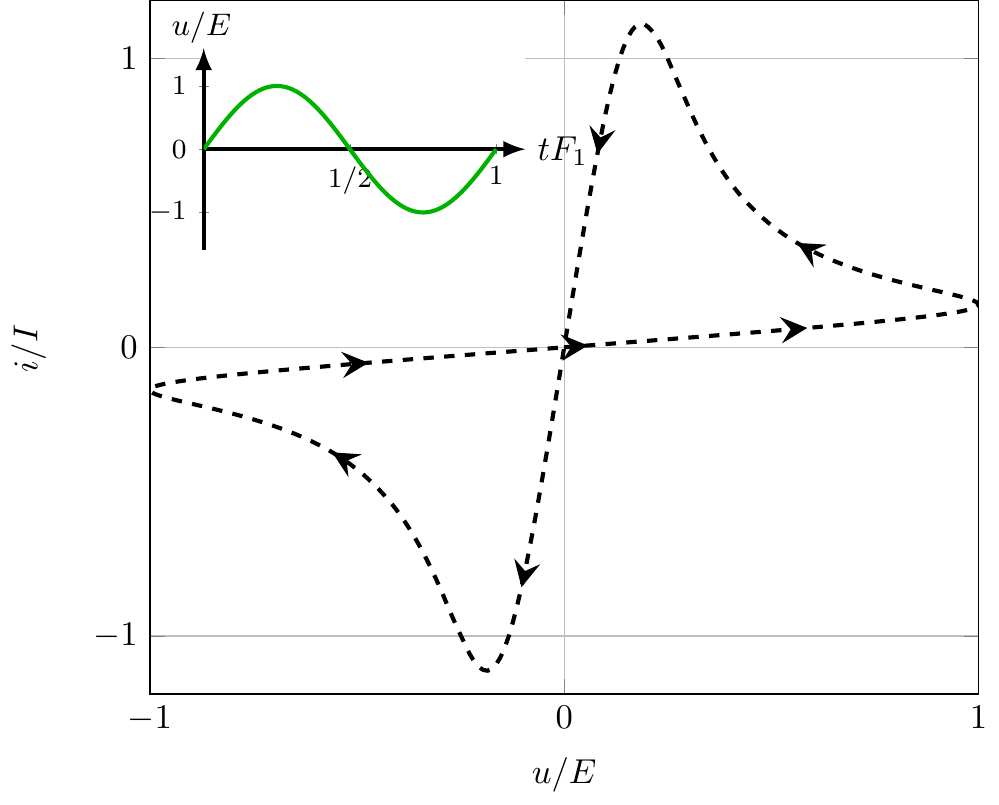}
	\caption{LTspice simulation of the HP-ion drift model.}
	\label{fig.:hpLTspiceHysteresisModel}
\end{figure}
} \solution{In this case, we have utilized a hysteresis curve generated by an LTspice implementation of the model~(\ref{eqn.:hpIonDriftModel}) in the mentioned identification and emulation procedure, as it can be seen in Fig.~\ref{fig.:hpLTspiceHysteresisModel}.
\begin{figure}[t!]
	\centering
	\includegraphics[scale=0.8]{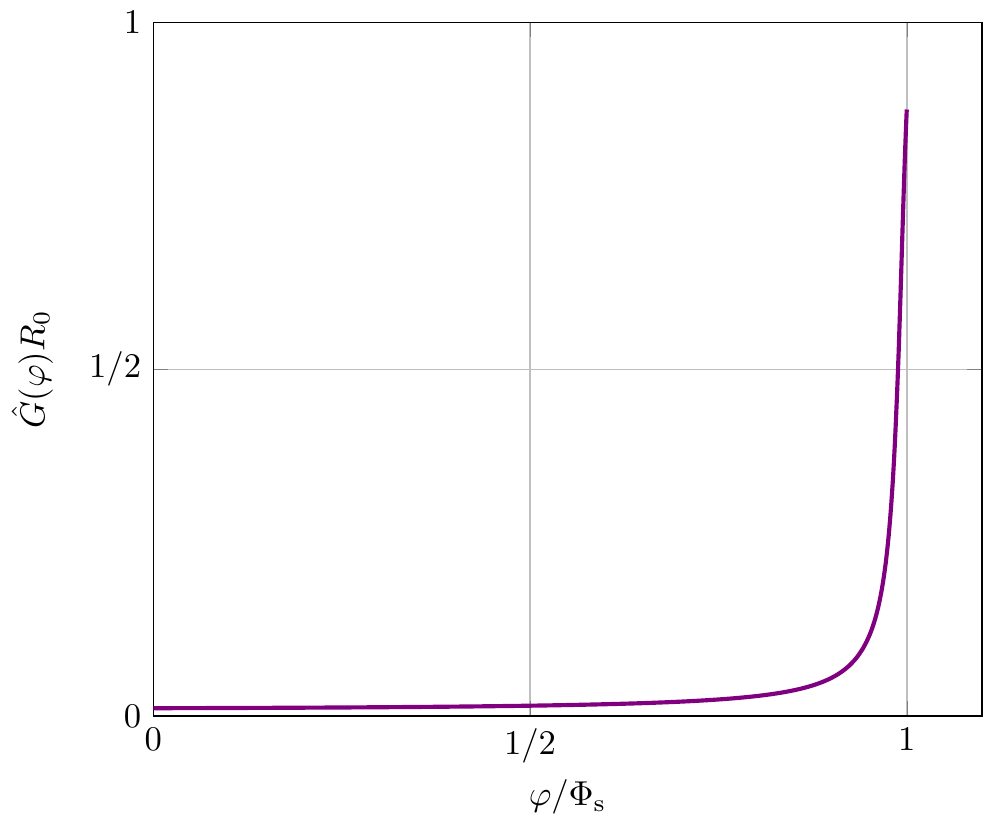}
	\caption{Characteristic curve of the HP-ion drift model generated from an LTspice simulation.}
	\label{fig.:hpIdentification}
\end{figure}
Identifying the memristor from simulated data yields the characteristic curve shown in Fig.~\ref{fig.:hpIdentification}. It should be mentioned that LTspice uses a variable step size for the numerical integration, which causes non-uniform samples for the voltage and current. From this information, we can uniformly resample the signals and process them as explained in Fig.~\ref{fig.:identification}.
\begin{figure}[t!]
	\centering
	\includegraphics[scale=0.8]{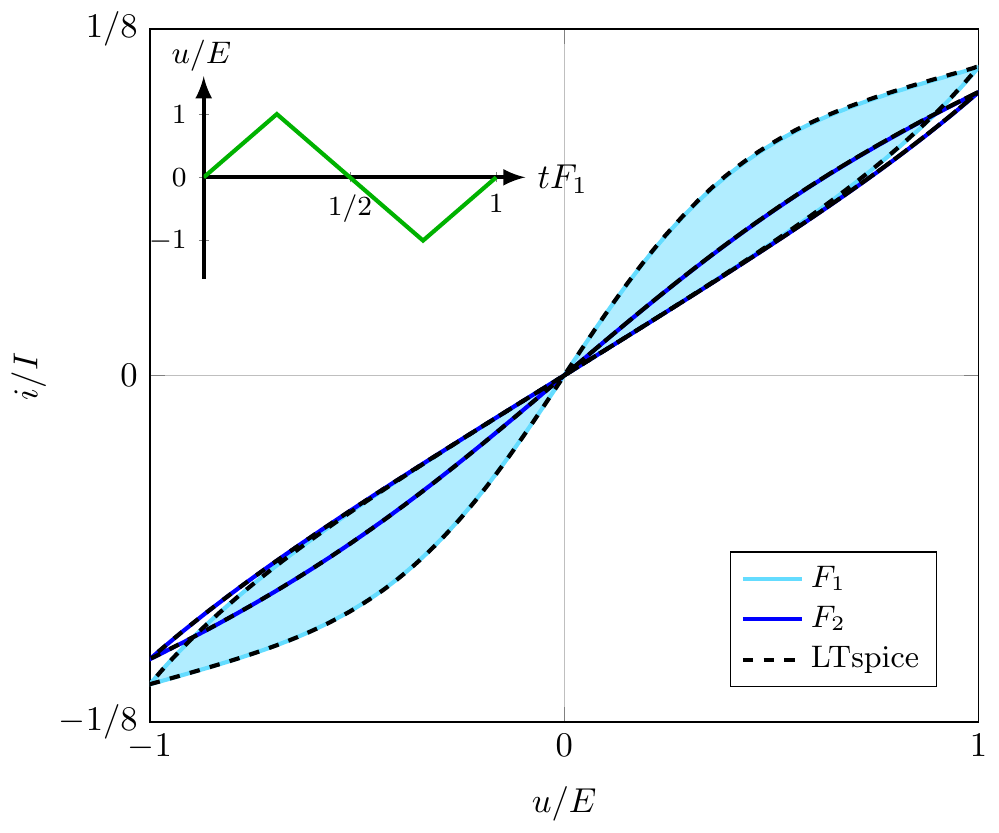}
	\caption{Wave digital validation of the identified HP-ion drift memristor.}
	\label{fig.:hpValidation}
\end{figure}
Validating the identified memristor by a triangular input signal regarding two distinct periods leads to the frequency-dependent hysteresis curves shown in Fig.~\ref{fig.:hpValidation}.  A confirmation of the identified device has been done by LTspice simulations (dashed curves) of the original model~(\ref{eqn.:hpIonDriftModel}). In order to identify the memristor in a broader range, other input signals, such as step functions, can be exploited.} \outlook{Note that, although the memristance and the constitutive relation are given in the model, we are not able to compute the flux-dependent memductance $\hat{G}(\varphi)$ explicitly. The Tab.~\RM{3} summarizes utilized parameters for this memristor.
\begin{table}[h!]
	\centering
	\renewcommand{\arraystretch}{1.3}
	\caption{Parameters HP-ion drift}
	\begin{tabular}{l r@{\,} r@{\,} l@{\,} l@{\,}}
		\hline
		\multicolumn{5}{c}{Software Emulation} \\
		Starting time        & $t_0$ & $=$ & $0$ & $\mathrm{s}$ \\
		Sampling period      & $T$   & $=$ & $1$ & $\mathrm{ms}$\\
		Number of iterations & $n_i$ & $=$ & $1$ &              \\
		\hline
		\hline
		\multicolumn{5}{c}{Excitation signals} \\
		Amplitude   & $E$   & $=$ & $1$ & $\mathrm{V}$ \\
		Frequency 1 & $F_1$ & $=$ & $1$ & $\mathrm{Hz}$\\
		Frequency 2 & $F_2$ & $=$ & $1,5$ & $\mathrm{Hz}$\\
		\hline
		\hline
		\multicolumn{5}{c}{Memristor} \\
		High resistance state & $R_1$ & $=$ & $10$   & $\mathrm{k\Omega}$   \\
		Low resistance state  & $R_0$ & $=$ & $100$ & $\mu\mathrm{\Omega}$\\
		Initial resistance    & $\hat{R}(z_0)$ & $=$ & $9$ & $\mathrm{k}\Omega$\\
		Material constant     & $\kappa$ & $=$ & $18,5$ & $[\mathrm{mC}]^{-1}$\\
		Exponent window function & $p$ & $=$ & $1$ & \\ 
		Normalization current  & $I$   & $=$ & $1$  & $\mathrm{mA}$   \\
		\hline
	\end{tabular}
\end{table}
}

It is remarkable that incorporating some kind of a characteristic function as a look-up table into the wave digital framework leads to very good emulation results even though no iteration for solving the implicit equation has been applied. Investigations regarding such implementations have shown that the naturally good conditioned system of equations originated by the wave digital algorithm is further improved by this approach.

\subsection{System-Model}
\label{ssec.:System-Model}%
\state{In many cases, a suited memristor for a dedicated application is desired. Therefore, we have to achieve devices with properties depending on the desired functionality of the overall system including them.} \aim{In this context, we are interested in a wave digital emulation of a particular voltage-controlled memristor (memductor) for a neuromorphic application.} \solution{For this reason, a generic model has been developed, which is incorporated into the wave digital algorithmic model.}
\subsubsection*{Multilevel Resistance Device}
\label{sssec.:Multilevel Resistance Device}
\state{In~\cite{ochs_anticipation_2017}, a memristive circuit mimicking the anticipation behavior of an amoeba has been extended to a circuit capable of anticipation of information represented by arbitrary digital patterns. There, a memristive device based on electrochemical metalization cells has been implemented. However, the wave digital implementation of the circuit with a more appropriate memristor model based on multilevel resistive switching devices has been presented in~\cite{ochs_wave_2017_02}. Actually, the closeness to reality makes these devices interesting for emulations~\cite{wang_sericin_2013,zhou_multilevel_2016}.} \aim{Our objective is to model a generic wave digital emulator of a multilevel memristor device. Depending on the particular application, the parameters of the model can be adjusted in order to get a device with a discrete up to a quasi-continuous resistance range.}

\solution{The proposed model
\begin{align}
	\begin{split}
		&i(t)
		= \hat{G}(z)\,u(t)\,,
		\quad \text{with}\quad
		\dot{z}(t)
		= u(t)+U_\mathrm{0}\,,\\
		&\hat{G}(z)
		= G_1-\frac{\Delta G}{n}h(z)\,,
		\,\,
		\Delta z
		= \frac{1}{n+1}\,,\,\,
		n\in\mathbb{N}\,,\\
		&\Delta G 
		= G_1-G_0\,,
		\quad 
		\sigma(\xi) 
		= \left\{ \begin{array}{rcl}
		1 & \mathrm{for} & \xi > 0\\
		0 & \mathrm{else} &
		\end{array} \right.\,,\\
		&\text{and}\quad
		h(z)
		=\sum_{\nu=1}^{n}\sigma(z-\nu\Delta z)+\sigma(-z-\nu\Delta z)
	\end{split}
\label{eqn.:multilevelModel}
\end{align}
includes a constant voltage $U_0$ in the memristive behavior for representing the retention characteristic of the device considering a more realistic emulator. Similar to the binary switching device, $G_1$ and $G_0$ denote the maximum and minimum conductance value, respectively. The conductance difference between $G_1$ and $G_0$ is abbreviated by $\Delta G$. In order to control the stepsize of memductance changing, a parameter $n$ is introduced, which also yields the stepsize for the state $\Delta z$. The expression $\sigma(\xi)$ denotes the unitstep function.}

\solution{The parameter $n$ leads to $n+1$ quantized memductance levels between $G_1$ and $G_0$, as it can be seen in the following investigations for $n=n_1=1$, $n=n_{10}=10$ and $n=n_{100}=100$.
\begin{figure*}[t!]
	\centering
	\subfloat[]{\includegraphics[scale=0.8]{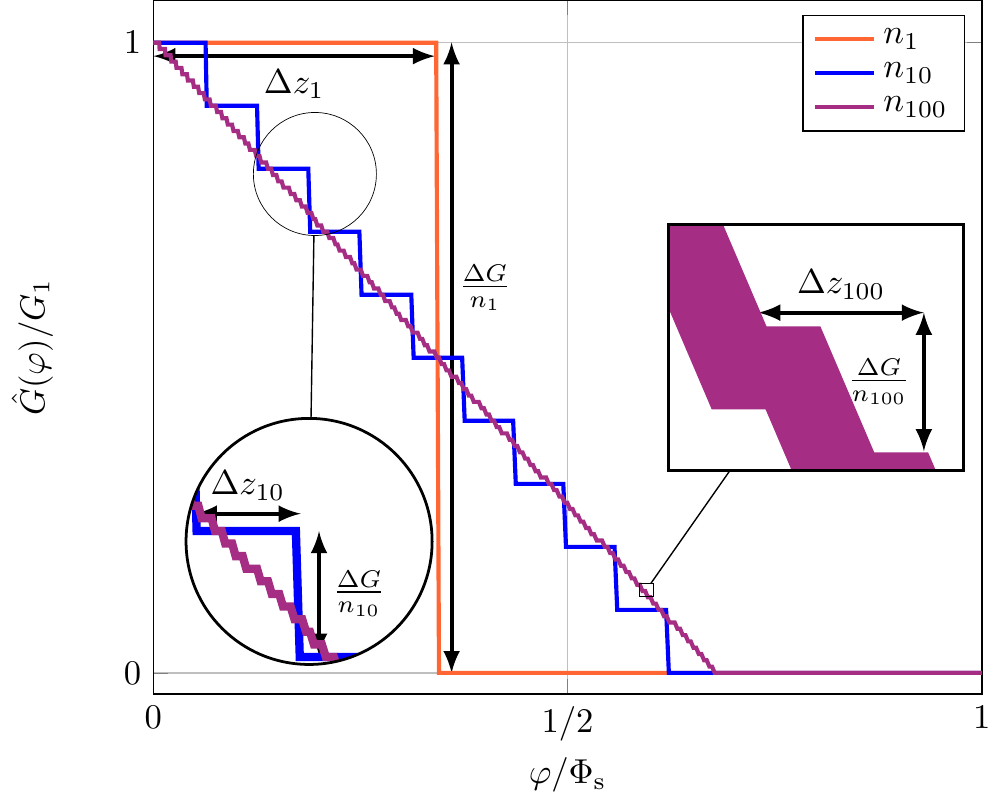}%
		\label{fig.:multilevelCharacteristic}}
	\hfil
	\subfloat[]{\includegraphics[scale=0.8]{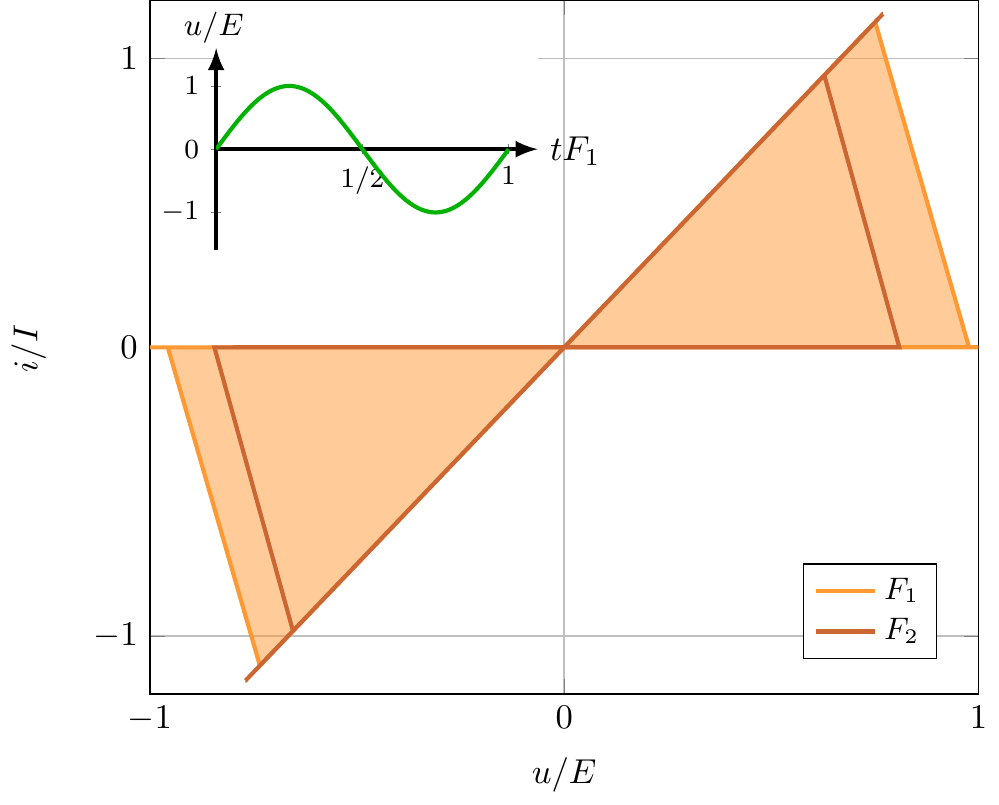}%
		\label{fig.:multilevelHysteresisModel_n1}}\\[0.2cm]
	\subfloat[]{\includegraphics[scale=0.8]{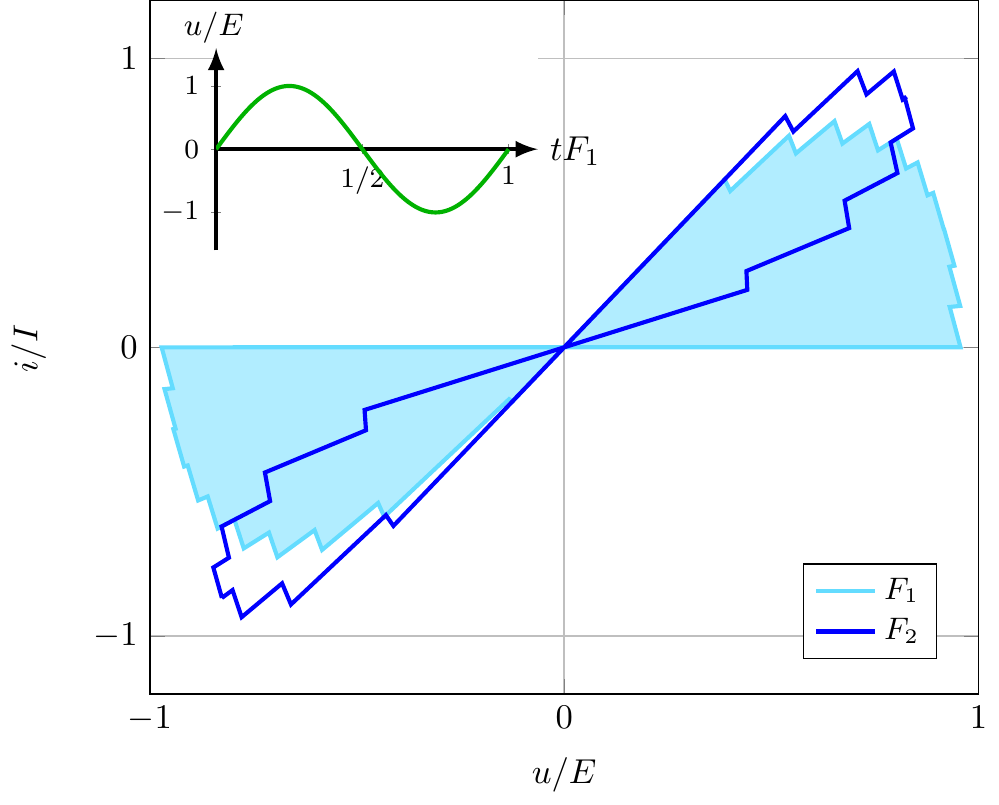}%
		\label{fig.:multilevelHysteresisModel_n10}}
	\hfil
	\subfloat[]{\includegraphics[scale=0.8]{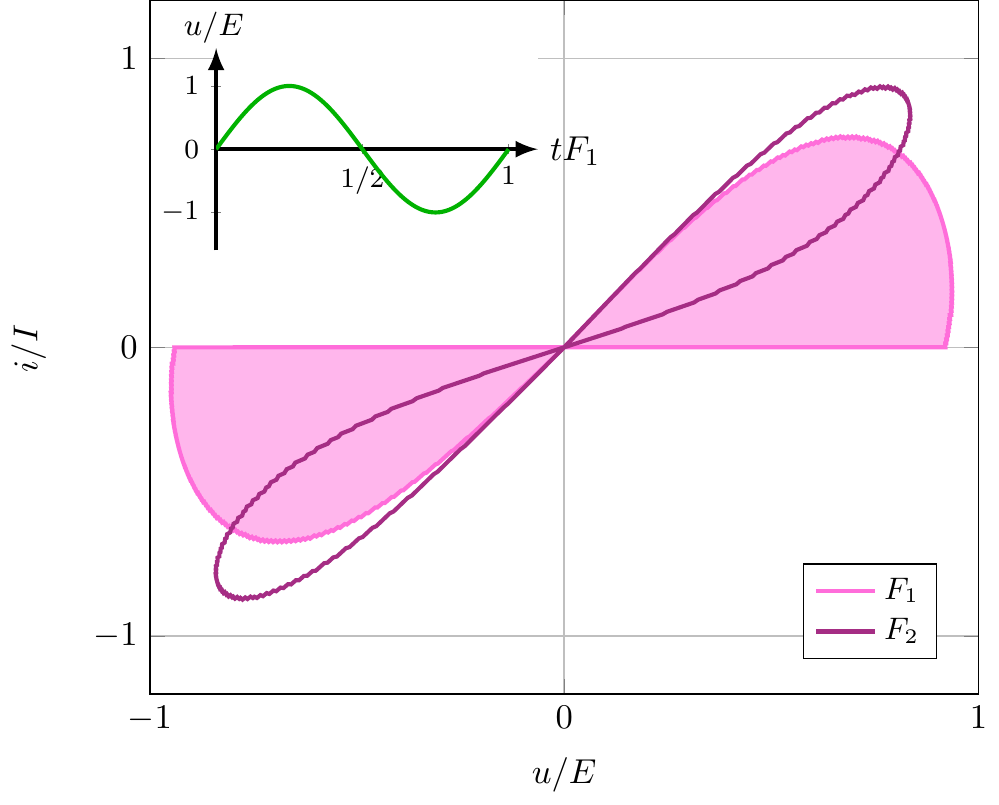}%
		\label{fig.:multilevelHysteresisModel_n100}}
	\caption{Characteristic curves of a multilevel resistance memristor (a) and resulting wave digital emulated hysteresis curves for $n_1=1$ (b), $n_{10}=10$ (c), and $n_{100}=100$ (d) refinement levels for two different excitation frequencies $F_1$ and $F_2$.}%
	\label{fig.:multilevel}%
\end{figure*}
The characteristic curves in Fig.~\ref{fig.:multilevelCharacteristic} indicate different memristor emulators, from a binary switch device for $n_1$ through a multilevel resistance memristor with $n_{10}=10$ distinct memductance levels, up to a quasi-continuous memristor for $n_{100}$, cf. Fig.~\ref{fig.:identificationBinarySwitch} and cf. Fig.~\ref{fig.:continuousHysteresisModel}, respectively. The resulting hysteresis curves of Fig.~\ref{fig.:multilevel} endorse this behavior. Similar to the before emulated models, the hysteresis area decreases for higher frequencies also for the multilevel memristor.} Utilized parameters can be observed from Tab.~\RM{4}.

\outlook{The proposed generic memristor model is suitable for different kinds of applications and can be adjusted with respect to the requirements in, e.g. nonvolatile memory applications (small $n$) or neuromorphic circuits (large $n$). Note that, with this model, we have combined a generic memristor model with a generic emulation technique, which makes our approach more flexible.} 

It is known that discontinuities could yield numerical problems, for example, considering LTspice implementations~\cite{biolek_reliable_2013}. Since the algorithmic model can also be utilized for simulation purposes, it is an appropriate method to investigate memristive devices with such discontinuities.

\begin{table}[h!]
	\centering
	\renewcommand{\arraystretch}{1.3}
	\caption{Parameters multilevel memristor}
	\begin{tabular}{l r@{\,} r@{\,} l@{\,} l@{\,}}
		\hline
		\multicolumn{5}{c}{Software Emulation} \\
		Starting time        & $t_0$ & $=$ & $0$ & $\mathrm{s}$ \\
		Sampling period      & $T$   & $=$ & $1$ & $\mathrm{ms}$\\
		Number of iterations & $n_i$ & $=$ & $1$ &              \\
		\hline
		\hline
		\multicolumn{5}{c}{Excitation signals} \\
		Amplitude   & $E$   & $=$ & $5$ & $\mathrm{V}$ \\
		Frequency 1 & $F_1$ & $=$ & $1$ & $\mathrm{Hz}$\\
		Frequency 2 & $F_2$ & $=$ & $2$ & $\mathrm{Hz}$\\
		\hline
		\hline
		\multicolumn{5}{c}{Memristor} \\
		High memductance state          & $G_1$ & $=$ & $3$   & $\mathrm{S}$   \\
		Low memductance state           & $G_0$ & $=$ & $100$ & $\mu\mathrm{S}$\\
		Reset voltage                   & $U_0$ & $=$ & $0,1$ & $\mathrm{V}$   \\
		Normalization current           & $I$   & $=$ & $10$  & $\mathrm{A}$   \\
		Low level refinement            & $n_1$ & $=$ & $100$ & 				  \\
		Intermediate level refinement   & $n_{100}$ & $=$ & $100$ & 				  \\
		High level refinement           & $n_{100}$ & $=$ & $100$ & 				  \\
		\hline
	\end{tabular}
\end{table}

\section{Emulation of Memristive Systems}
\label{sec.:Emulation of Memristive Systems}
\state{In contrast to memristors, general memristive systems can have more than one internal state. In context of memristive systems, the internal states are not necessarily the electrical charge or the magnetic flux. The most often realized devices can be described by a general memristive system~(\ref{eqn:memristiveSystemElectrical}).} \aim{In order to emulate such devices, a physical model of the memristive system, which can be transformed into a reference circuit regarding a wave digital emulation, is desired.} \solution{In the following, the wave digital emulation of a double barrier memristive device (DBMD) based on a physical model of the real device is presented.}

\subsection*{Physical Model}
\label{ssec.:Physical Model}%
\state{In order to emulate real memristive devices, a reference circuit of the physical model, which can be transformed into the wave digital domain, is required. 
\begin{figure}[t!]
	\centering
	\includegraphics[scale=0.6]{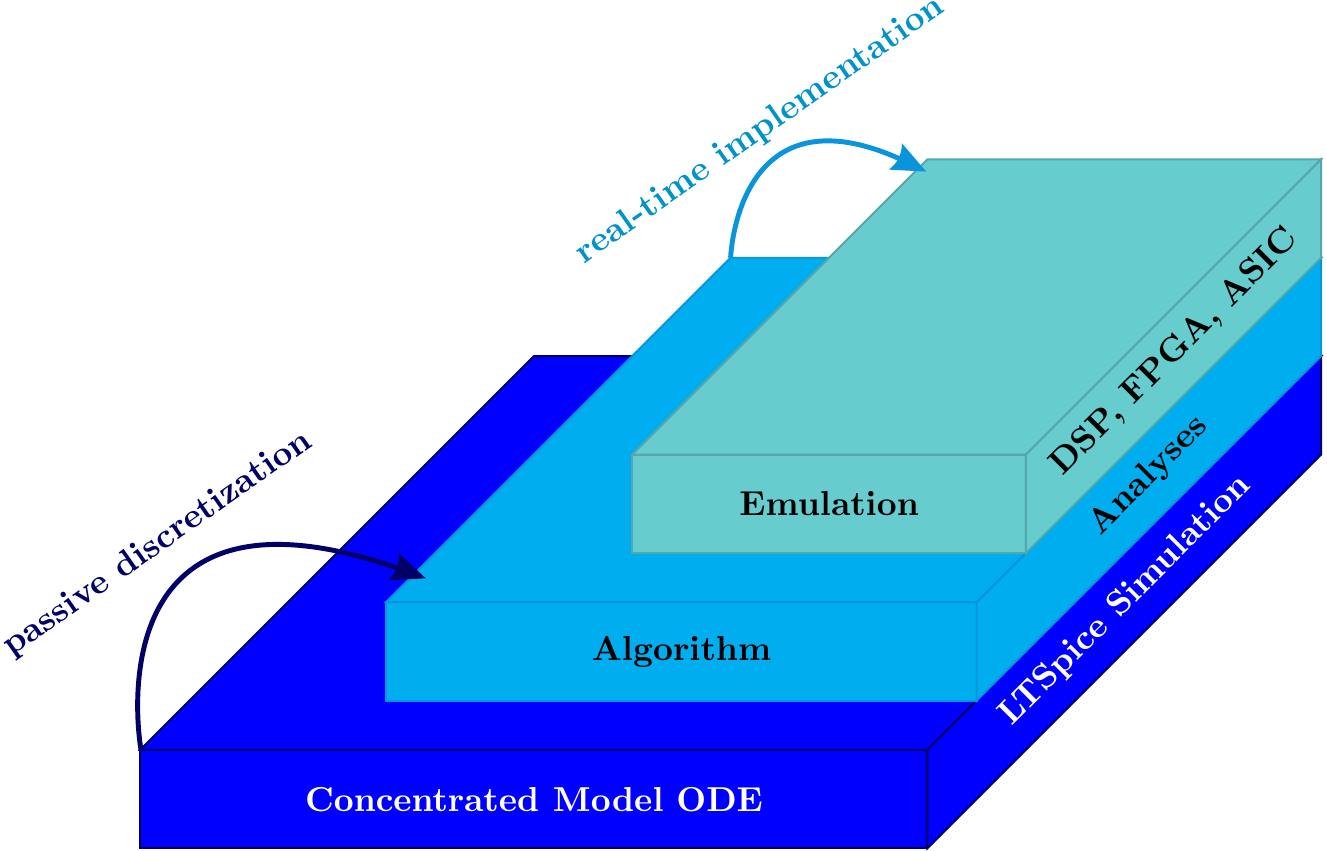}%
	\caption{Stages of development toward an emulator of a physical device.}%
	\label{fig.:procedure2}%
\end{figure}
Once the reference circuit, also called equivalent circuit or model with lumped/concentrated parameters, is determined, it can be utilized to get an algorithmic model, which in turn can be implemented platform-independent, cf. Fig~\ref{fig.:procedure2}.} \solution{Based on this workflow, the wave digital method has been applied to a concentrated model of the double barrier memristive device for an emulation on a digital signal processing unit.}

\subsubsection*{Double Barrier Memristive Device}
\label{sssec.:Double Barrier Memristive Device}%
\state{The DBMD is a manufactured memristive system with beneficial properties especially in neuromorphic applications, like a quasi-continuous resistance range, no need for an electric forming procedure, and an improved retention characteristic~\cite{hansen_double_2015}. Investigations on an atomic level based on a model with distributed parameters in kinetic Monte-Carlo simulations can be seen in~\cite{dirkmann_role_2016}.} \causes{Although a distributed model is suitable for identifying the underlying physical and chemical phenomena within the device, it is less convenient for a real-time capable implementation.} \aim{A wave digital realization of this device can overcome this problem.} \solution{As mentioned before, a concentrated model, called reference circuit, is needed in order to transform it into the wave digital domain. The whole modeling procedure of the device can be seen in~\cite{solan_enhanced_2017} and is briefly recapitulated here: 
\begin{figure}[hbt!]
	\centering
	\includegraphics[scale=0.9]{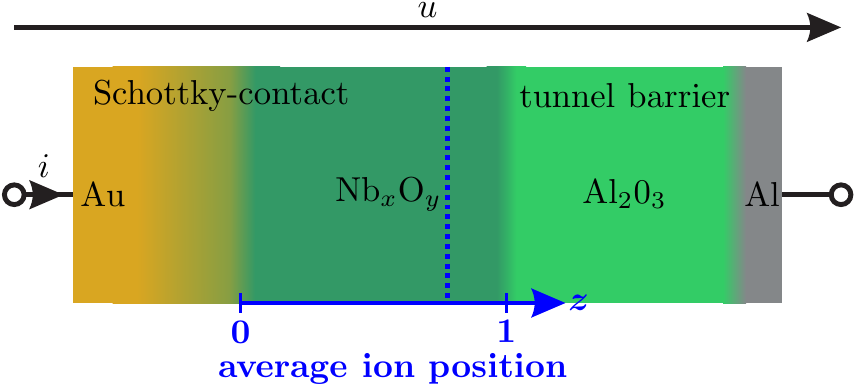}\\[1ex]%
	\includegraphics[scale=0.9]{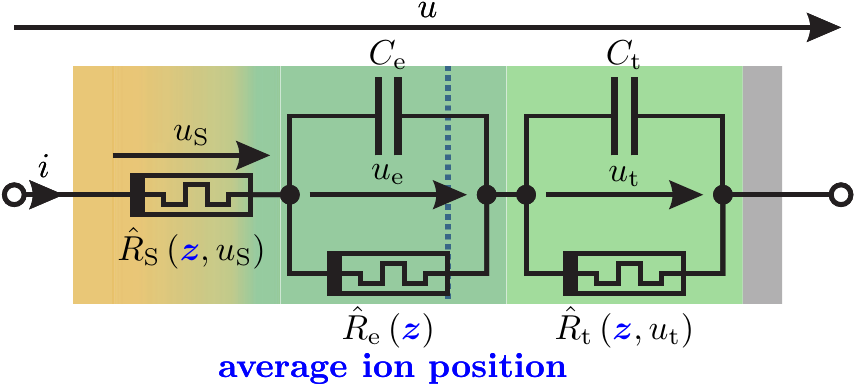}
	\caption{Material composition of the real device and corresponding equivalent circuit.}%
	\label{fig.:electricalModel}%
\end{figure}
A lumped element electrical model with concentrated but still physically meaningful parameters is shown in Fig.~\ref{fig.:electricalModel}. The DBMD mainly consists of a solid-state electrolyte (niobium oxide) next to a tunnel barrier region (aluminum oxide) sandwiched between two electrodes (gold and aluminum). Indeed, the metal-electrolyte interface builds physically a Schottky-like contact, which is state- and voltage-dependent.  Due to the fact, that the average ion position within the solid-state electrolyte modulates the interface effects between distinguish regions, it has been identified as the internal state of the memristive system. The memristive behavior of the device
\begin{align}
	\begin{split}
		&\dot{z}
		= \frac{-\dot{Z}\,w(z)}{e^{\varphi_\mathrm{a}(u,z)}}\sinh\left(\frac{\,u_\mathrm{r}(u,u_\mathrm{S},z)+u_\mathrm{e}-U_\mathrm{c}}{U_\mathrm{e}}\right)\:,\\
		&\text{with}\quad
		w(z)
		= \left[1-2\,w_0\right]\left[1-\left[2\,z-1\right]^{2p}\right]+w_0
	\end{split}
\end{align}
is based on an ion hopping phenomenon. In order to overcome the boundary lock problem, the window function of the HP-ion drift model has been extended to the proposed window function $w(z)$, where $w_0>0$ is a constant to prevent that the window function becomes zero, cf.~\cite{solan_enhanced_2017}. The normalized average ion velocity $\dot{Z}$ and the reference electrolyte voltage $U_\mathrm{e}$ as well as the Coulomb voltage $U_\mathrm{C}$ are electrical constants with a physical interpretation. The ion motion within the electrolyte depends mainly on the voltage drop over this region $u_\mathrm{e}$. However, from physical measurements as well as simulations with the distributed model on an atomic scale, we know that the ions next to the Schottky-interface are also influenced by the voltage over this interface. As a consequence a voltage- and state-dependent amount of this voltage which is described by 
\begin{align}
	u_\mathrm{r}(u,u_\mathrm{S},z) 
	&= \sigma(-u)\left[1-z\right]\,u_\mathrm{S}\:
	\label{eqn:schottkyVoltageAmount}
\end{align}
must be considered for the ion motion. Here, $\sigma(\xi)$ is again the unitstep function, cf. equation~(\ref{eqn.:multilevelModel}), whereas $u_\mathrm{S}$ denotes the voltage over the Schottky-contact. A particular voltage- and state-dependent normalized activation energy captures some novel physical insights, like the adsorption and desorption mechanisms of ions at the interfaces. This describes the mobility of ions within the electrolyte and has been introduced by
\begin{equation}
\varphi_\mathrm{a}(u,z) = \sigma(u)\left[\varphi_{\mathrm{a}_1}+z\left[\varphi_{\mathrm{a}_0}-\varphi_{\mathrm{a}_1}\right]-\varphi_{\mathrm{a_r}}\right]+\varphi_{\mathrm{a_r}}\:,
\label{eqn:stateDependentActivationEnergy}
\end{equation}
where $\varphi_{\mathrm{a}_0}$ and $\varphi_{\mathrm{a}_1}$ correspond to the state $z=0$ and $z=1$, respectively. The remaining parameter $\varphi_{\mathrm{a_r}}$ is the normalized activation energy valid for the reverse direction $u\leq 0$ and leads to the typical retention characteristic.}

\solution{Beyond the memristive behavior, the input-output relations are also needed for representing memristive systems, cf. equation~(\ref{eqn:memristiveSystemElectrical}). Since the device includes three memristive systems each of them coupled via the same internal state, it can be segmented into three input-output relations, see Fig.~\ref{fig.:electricalModel}. The first one is described by the Schottky-equation representing the voltage- and state-dependent Schottky-contact in the concentrated model
\begin{align}
	\hat{R}_\mathrm{S}(u_\mathrm{S},z)
	&= \frac{u_\mathrm{S}}{I_\mathrm{S}}\frac{e^{\left[\varphi_\mathrm{S}(z)+\alpha_\mathrm{f}\sqrt{\frac{|u_\mathrm{S}|-u_\mathrm{S}}{\alpha_\mathrm{S}\,U_\vartheta}}\right]}}{e^{\frac{1}{n(z)}\frac{u_\mathrm{S}}{U_\vartheta}}-1}\:.
\end{align}
Here, a dimensionless fitting parameter $\alpha_\mathrm{f}$ weights the barrier lowering term, which is caused by the Schottky-effect. The parameter $\alpha_\mathrm{S}$ is the normalized Schottky-barrier thickness and $I_\mathrm{S}$ denotes the Schottky-current amplitude including some physical constants and material parameters, cf.~\cite{solan_enhanced_2017}. $U_\vartheta$ is the thermal voltage caused by the temperature. From physical arguments, a state-dependent Schottky-barrier height as well as ideality factor have been identified
\begin{align}
	\begin{split}
		\varphi_\mathrm{s}(z) 
		&= \varphi_{\mathrm{s}_0}+z\left[\varphi_{\mathrm{s}_1}-\varphi_{\mathrm{s}_0}\right]
		\quad\mathrm{and}\\
		n(z) 
		&= n_0+z\left[n_1-n_0\right]\:,
	\end{split}
\end{align}
with $\varphi_{\mathrm{s}_0}$/$\varphi_{\mathrm{s}_1}$ describing the lower/upper bounds of the Schottky-barrier height and $n_0$/$n_1$ representing the lowest/highest limit of the ideality factor in the Schottky-equation.}

\solution{The second partial memristive system is the electrolyte region itself. For this region, a linear change of the resistance between the high $R_{\mathrm{e}_1}$ and low resistance $R_{\mathrm{e}0}$ states is assumed
\begin{align}
	\hat{R}_\mathrm{e}(z)
	&= R_{\mathrm{e}_0}+z\left[R_{\mathrm{e}_1}-R_{\mathrm{e}_0}\right]\:.
\end{align}
}

\solution{The outstanding memristive system is caused by a state- and voltage-dependent tunnel barrier region. Based on the very fundamental Simmon's equation~\cite{simmons_generalized_1963}, the memristance of this region can be described by
\begin{align}
	\begin{split}
		\hat{R}_\mathrm{t}(u_\mathrm{t},z)
		&= \frac{u_\mathrm{t}}{I_\mathrm{t}}\frac{\alpha_\mathrm{t}^2(z)}{g(-u_\mathrm{t},z)-g(u_\mathrm{t},z)}\:,
		\quad\text{with}\\
		g(u_\mathrm{t},z)
		&= \varphi_\mathrm{t}(u_\mathrm{t})e^{-\alpha_\mathrm{t}(z)\sqrt{\varphi_\mathrm{t}(u_\mathrm{t})}}
		\quad\text{and}\\
		\varphi_\mathrm{t}(u_\mathrm{t})
		&= \varphi_{\mathrm{t}_0}+\frac{u_\mathrm{t}}{2\,U_\vartheta}\:.
	\end{split}
\end{align}
Here, $\varphi_{\mathrm{t}_0}$ is a normalized tunnel barrier thickness utilized as reference and $u_\mathrm{t}$ stands for the voltage drop over the tunnel barrier. In order to make the region memristive, a state-dependent normalized tunnel barrier thickness is assumed
\begin{align}
	\alpha_\mathrm{t}(z) 
	&= \alpha_{\mathrm{t}_0}+z\left[\alpha_{\mathrm{t}_1}-\alpha_{\mathrm{t}_0}\right]\:,
\end{align}
where $\alpha_{\mathrm{t}_0}=\alpha_\mathrm{t}(z=0)$ and $\alpha_{\mathrm{t}_1}=\alpha_\mathrm{t}(z=1)$.} \outlook{A detailed physical interpretation of the proposed concentrated model can be caught up from~\cite{solan_enhanced_2017}.}

\causes{Although the concentrated model is suitable for pre-investigations, e.g. by an LTspice implementation, it is still not capable for real-time emulations. Note that a simulation model cannot replace real devices in real circuit as emulators.} \aim{We intend a flexible emulator of the DBMD in order to make investigations with respect to different parameters.} \solution{We have already shown that the wave digital method is a very effective approach to achieve such an emulator. First investigations with a wave digital realization of a reduced model of the DBMD have been done recently~\cite{ochs_wave_2016}. Here, we introduce the complete model of the DBMD, which is used as a reference circuit for the wave digital realization of the device.
\begin{figure*}[t!]
	\centering
	\includegraphics[scale=0.9]{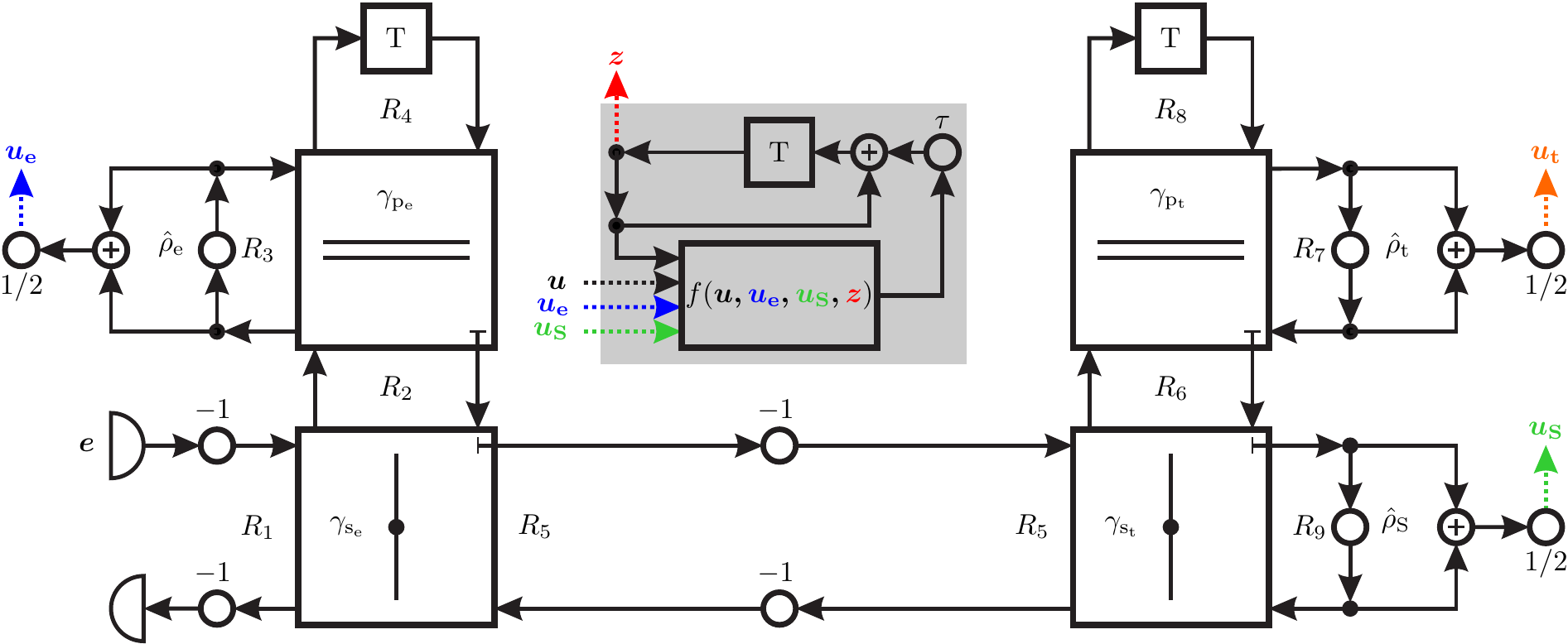}%
	\caption{Wave flow diagram of the double barrier memristive device, with $u=u_\mathrm{S}+u_\mathrm{e}+u_\mathrm{t}$.}%
	\label{fig.:waveDigitalModel}%
\end{figure*}
The resulting wave flow diagram is depicted in Fig.~\ref{fig.:waveDigitalModel}. There, the Kirchhoff's parallel and series interconnections are represented by so called series and parallel adaptors with corresponding adaptor coefficients 
\begin{align}
	\gamma_\mathrm{s}
	&= \frac{R_1}{R_3}
	\quad\text{and}\quad
	\gamma_\mathrm{p}
	= \frac{R_2}{R_3}\:,
\end{align}
with $R_1$ and $R_2$ as the port resistance at port 1 and 2, respectively. Note that the port 3 is regarded as a reflection free port, thus the port resistance at this port can be determined by $R_3=R_1+R_2$. Signal flow diagrams of such elements can be seen in~\cite{fettweis_wave_1986} and are not the focus of this work. Similar to the wave digital realization of the integrator circuit of Fig.~\ref{fig.:ElectricalInterpretationMemory}, it can be seen that the parasitic capacitors are simply realized as delay elements, whereas voltage- and state-dependent reflection coefficients indicate a memristive system, cf. equation~(\ref{eqn.:memristiveSystemWd}). The port resistances are denoted by $R_\nu\,,\:\nu=1,2,...,9$ being positive values. For the capacitive port they can be determined as $R_4=T/(2C_\mathrm{e})$ and $R_8=T/(2C_\mathrm{t})$, cf. with the integrator circuit in Fig.~\ref{fig.:ElectricalInterpretationMemory}. As it turned out, the port resistances influence the conditioning of the equation system, which has to be solved. This means that, for a convenient choice of the port resistance values, the algorithm robustness can be increased. The Tab.~\RM{5} at the end of the section summarizes concrete values of all utilized parameters. Corresponding to Fig.~\ref{fig.:waveDigitalRealizationMemristiveDevice}, the pre-processing unit including the memristive function $f$ and the integrator, which determines the actual internal state, is emphasized by a gray background color in Fig.~\ref{fig.:waveDigitalModel} representing the wave flow diagram. It is notable that the memristive function in the algorithm can arbitrarily be predefined in order to achieve different memristive behaviors. This underlies the flexibility of the proposed emulator even for emulating real devices. The emulation of the double barrier memristive device also indicates that the proposed approach is also suitable for emulation whole sub-circuits instead of a single device. 
\begin{figure}[t!]
	\centering
	\includegraphics[scale=0.8]{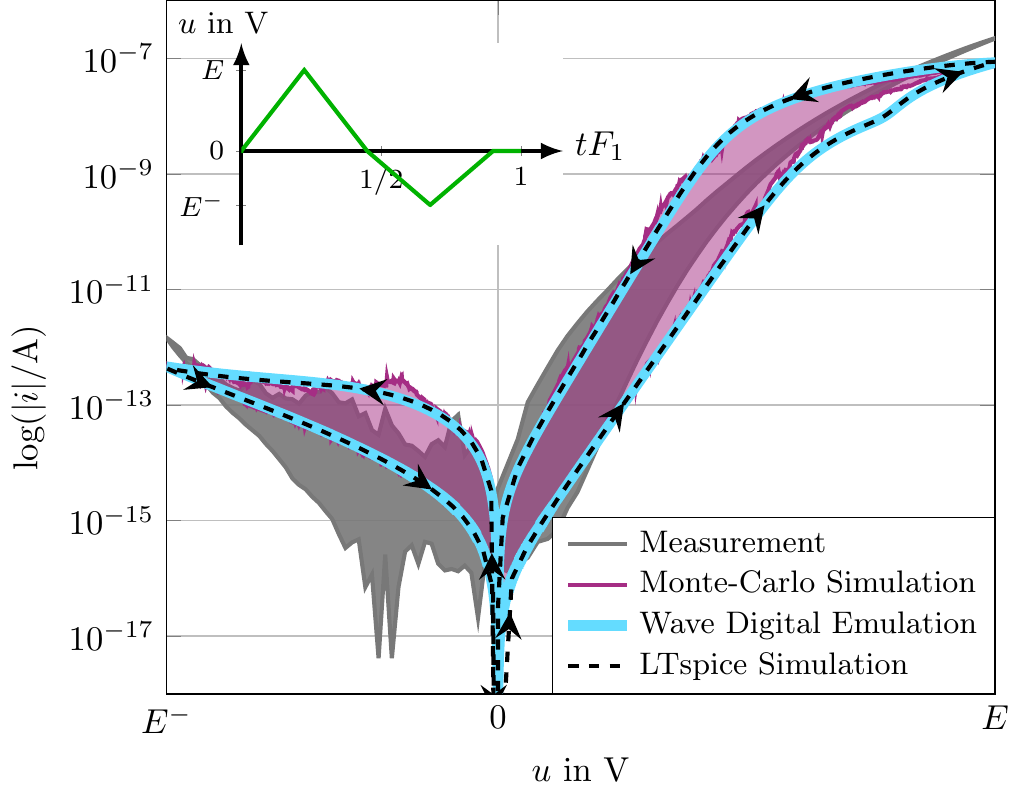}
	\caption{Wave digital emulated hysteresis curve of the DBMD compared to a measurement of the real device, kinetic Monte-Carlo simulation of the model with distributed parameters on an atomic level and an LTspice implementation of the reference circuit.}
	\label{fig.:hysteresisWD}
\end{figure}
The resulting hysteresis curve considering a triangular input voltage is shown in Fig.~\ref{fig.:hysteresisWD}. There, the wave digital emulation is compared to the measurement of the real device, to a kinetic Monte-Carlo simulation of the model with distributed parameters, and to an LTspice simulation of the reference circuit. A good coincidence with the measurement as well as kinetic Monte-Carlo simulation can be observed. In addition, there is an almost perfect coincidence between the wave digital emulation and the LTspice simulation. This indicates a correct transformation of the concentrated model into the wave digital domain.}

\solution{Investigations with respect to the frequency dependency of the device has been done by a sinusoidal input signal, cf.~(\ref{eqn.:sinusoidalInput}).
\begin{figure}[t!]
	\centering
	\includegraphics[scale=0.8]{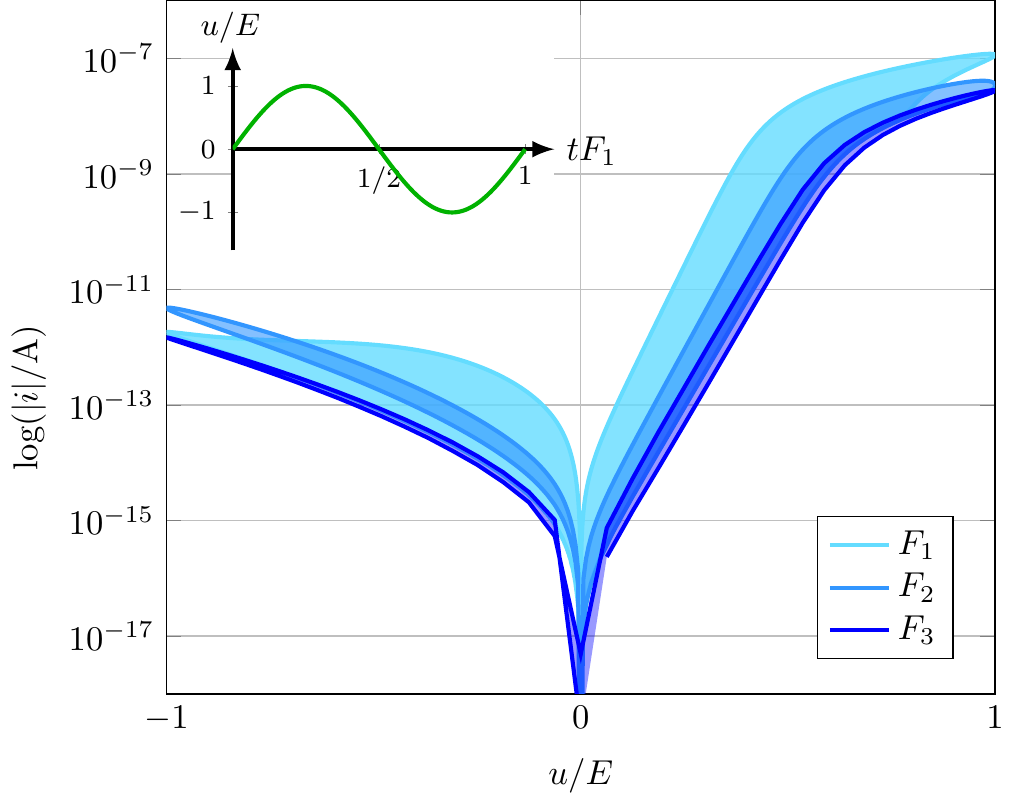}
	\caption{Wave digital emulated hysteresis curve of the DBMD for a sinusoidal excitation with three different frequencies.}
	\label{fig.:frequencyDependenceHysteresis}
\end{figure}
The results of the frequency-dependent hysteresis curve are depicted in Fig.~\ref{fig.:frequencyDependenceHysteresis}, which shows again a decreasing hysteresis area for higher frequencies.}

\outlook{The proposed approach can also be utilized for further investigations on a simulation level. Several benefits distinguish simulations based on a wave digital algorithm from usual simulation tools. For example, the fact that the algorithmic model only consists of elementary mathematical operations supports a platform-independent implementation. In addition, the wave digital algorithm is based on a passive electrical reference circuit, which inherently leads to well-conditioned systems of equations. This, in turn, results in very efficient and robust algorithms. For example, the simulation of the shown hysteresis curve in Fig.~\ref{fig.:hysteresisWD} takes several days up to a week when applying a model with distributed parameters, whereas a wave digital algorithm needs seconds. Even the measurement of the real device takes longer than the wave digital algorithm and hence it is especially suitable for investigations of the device even before fabrication. With this, the development cycles of novel innovative devices can be decreased. Moreover, the algorithm can be utilized for further signal processing approaches, e.g. in~\cite{solan_parameter_2017} a model reduction of the DBMD by maintaining the functionality of the real device has been shown. For this purpose, methods known from optimization theory have been applied to the wave digital algorithm in order to find optimal parameters for the reduced model. This is quite interesting because this powerful tool in combination with physically meaningful parameters in the model can be used for some kind of reverse engineering. Indeed, we can achieve desired functionalities by optimizing the parameters. A reinterpretation of the obtained parameters, for example, could result in novel material compositions, different physical dimensions or particular doping concentrations. The material scientists can be served with this information and thus we can make investigations on a material or physical level in order to develop novel memristive devices even before fabrication. Particularly, the wave digital model of the double barrier memristive device has been exploited in order to find physical and chemical effects of the real device, which were not known before, like adsorption and desorption of ions at the interfaces. These effects have significant influences on the long time scale behavior, which has been improved by the proposed method. All of these aspects are of great importance especially in terms of the development costs of such devices.}

\outlook{Note that we are not constrained to emulation of memristive devices with this method. Instead of a single device the wave digital emulation of the DBMD has shown that complete sub-circuits can also be emulated. In this context, influence investigations considering well-defined parameter spread in complex systems including such devices can be done in a very effective manner. Moreover, memreactive systems, thus capacitances or inductances with memory are appropriate candidates for emulations by the proposed method because fabrication of such elements is even harder than memristive devices. A consistent modeling of memcapacitive devices, which can be utilized as a reference circuit, is shown in~\cite{ochs_consistent_2017}.}

\begin{table}[h!]
	\centering
	\renewcommand{\arraystretch}{1.3}
	\caption{Parameters double barrier memristive device}
	\begin{tabular}{l r@{\,} r@{\,} l@{\,} l@{\,}}
		\hline
		\multicolumn{5}{c}{Software Emulation} \\
		Starting time        & $t_0$ & $=$ & $0$ & $\mathrm{s}$ \\
		Sampling period      & $T$   & $=$ & $10$ & $\mathrm{ms}$\\
		Number of iterations & $n_i$ & $=$ & $6$ &              \\
		\hline
		\hline
		\multicolumn{5}{c}{Excitation signals} \\
		Positive amplitude   & $E$   & $=$ & $3$ & $\mathrm{V}$ \\
		Negative amplitude   & $E^-$   & $=$ & $-2$ & $\mathrm{V}$ \\
		Frequency 1 & $F_1$ & $=$ & $10$ & $\mathrm{mHz}$\\
		Frequency 2 & $F_2$ & $=$ & $100$ & $\mathrm{mHz}$\\
		Frequency 3 & $F_3$ & $=$ & $1$ & $\mathrm{Hz}$\\
		\hline
		\hline
		\multicolumn{5}{c}{Memristive Behavior} \\
		Normalized average ion velocity         & $\dot{Z}$                & $=$&$0.32$           & $\mathrm{THz}$\\
		Reference electrolyte voltage           & $U_\mathrm{e}$           & $=$&$323.2$          & $\mathrm{mV}$\\
		Minimum normalized activation energy    & $\varphi_{\mathrm{a_0}}$ & $=$&$26.3$           & \\
		Maximum normalized activation energy    & $\varphi_{\mathrm{a_1}}$ & $=$&$36.75$          & \\
		Normalized reset activation energy      & $\varphi_{\mathrm{a_r}}$ & $=$&$30.17$          & \\
		Window function offset                  & $w_0$                    & $=$&$100$            & $\mu$  \\
		Window function exponent                & $p$                      & $=$&$6$              & \\
		Coulomb voltage                         & $U_\mathrm{c}$           & $=$&$0.1$            & $\mathrm{mV}$\\
		\hline
		\hline
		\multicolumn{5}{c}{Electrolyte} \\
		Minimum electrolyte resistance          & $R_{\mathrm{e_0}}$       & $=$&$2$              & $\mathrm{M}\Omega$\\
		Maximum electrolyte resistance          & $R_{\mathrm{e_1}}$       & $=$&$5.1$            & $\mathrm{M}\Omega$\\
		Electrolyte capacitance                 & $C_\mathrm{e}$           & $=$&$17.4$           & $\mathrm{fF}$ \\
		\hline
		\hline
		\multicolumn{5}{c}{Schottky-Contact} \\
		Normalized min Schottky-barrier height & $\varphi_{\mathrm{s_0}}$ & $=$&$27.08$ & \\
		Normalized max Schottky-barrier height & $\varphi_{\mathrm{s_1}}$ & $=$&$34.81$ & \\
		Normalization Schottky-barrier thickness   & $D_\mathrm{s}$           & $=$&$1.326$ & $\mathrm{nm}$\\
		Normalized Schottky-barrier thickness      & $\alpha_{\mathrm{s}}$    & $=$&$3.77$  & \\
		Schottky current amplitude                 & $I_\mathrm{s}$           & $=$&$108$   & $\mathrm{mA}$\\
		Minimum ideality factor         & $n_0$                 & $=$&$2.9$   & \\
		Maximum ideality factor         & $n_1$                 & $=$&$4.1$   & \\
		Fitting parameter                          & $\alpha_\mathrm{f}$      & $=$&$-1.25$ & \\
		Thermal voltage                           & $U_\vartheta$ & $=$&$26$& $\mathrm{mV}$\\
		\hline
		\hline
		\multicolumn{5}{c}{Tunnel-Barrier} \\
		Normalized tunnel barrier height            & $\varphi_{\mathrm{t_0}}$ & $=$&$108.32$ & \\
		Normalized min tunnel barrier thickness & $\alpha_{\mathrm{t_0}}$  & $=$&$1.81$   & \\
		Normalized max tunnel barrier thickness & $\alpha_{\mathrm{t_1}}$  & $=$&$2.03$   & \\
		Tunnel barrier current amplitude            & $I_\mathrm{t}$           & $=$&$432.6$  & $\mathrm{mA}$\\
		Tunnel barrier capacitance                  & $C_\mathrm{t}$           & $=$&$20.7$   & $\mathrm{fF}$ \\
		\hline
		\hline
		\multicolumn{5}{c}{Wave Digital Parameters (freely selectable port resistances)} \\
		Port resistance 1            & $R_1$ & $=$ & $1$ & $\Omega$\\
		Port resistance 3            & $R_3$ & $=$ & $10$ & $\mathrm{M}\Omega$\\
		Port resistance 7            & $R_7$ & $=$ & $1$ & $\mathrm{G}\Omega$\\
		\hline
	\end{tabular}
\end{table}

\section{Conclusion}
In the present paper, a generic wave digital emulation of memristive devices has been presented. Generic in this context means that the memristive model itself, as well as the concrete realization of the emulator, is not restricted to a particular choice. A generic model based on the multilevel resistance device for applications in nonvolatile memory but also in neuromorphic circuits has been proposed. Additionally, a special identification procedure has been exploited in order to emulate devices with respect to a desired functionality given by the hysteresis curve. Additionally, it has been also shown that a physically more accurate modeling approach can be utilized in order to get emulators of real devices, like the double barrier memristive device.

Several benefits of the wave digital approach for emulating memristive devices have been figured out. Especially the use of the algorithmic model in optimization procedures in combination with physically meaningful parameters make this approach very useful for the development of innovative memristive device. Since the presented method can also be used to get an algorithmic model of whole sub-circuits instead of single devices, efficient influence investigations of well-defined parameter spread with respect to the functionality of the single device as well as to the overall functionality of more complex neuromorphic circuits can be done. In fact, these investigations are reproducible in contrast to investigations with real devices and they are even faster than real measurements.  

\section*{Acknowledgment}
The financial support by the German Research Foundation (Deutsche Forschungsgemeinschaft - DFG) through FOR 2093 is gratefully acknowledged. We thank also our colleagues from the research group FOR 2093 Mirko Hansen who provided measured data of the DBMD and Sven Dirkmann for his kinetic Monte-Carlo simulation results.

\bibliographystyle{IEEEtran}
\bibliography{IEEEabrv,bibliography}

\begin{thebibliography}{10}
\providecommand{\url}[1]{#1}
\csname url@samestyle\endcsname
\providecommand{\newblock}{\relax}
\providecommand{\bibinfo}[2]{#2}
\providecommand{\BIBentrySTDinterwordspacing}{\spaceskip=0pt\relax}
\providecommand{\BIBentryALTinterwordstretchfactor}{4}
\providecommand{\BIBentryALTinterwordspacing}{\spaceskip=\fontdimen2\font plus
\BIBentryALTinterwordstretchfactor\fontdimen3\font minus
  \fontdimen4\font\relax}
\providecommand{\BIBforeignlanguage}[2]{{%
\expandafter\ifx\csname l@#1\endcsname\relax
\typeout{** WARNING: IEEEtran.bst: No hyphenation pattern has been}%
\typeout{** loaded for the language `#1'. Using the pattern for}%
\typeout{** the default language instead.}%
\else
\language=\csname l@#1\endcsname
\fi
#2}}
\providecommand{\BIBdecl}{\relax}
\BIBdecl

\bibitem{indiveri_memory_2015}
G.~Indiveri and S.-C. Liu, ``Memory and information processing in neuromorphic
  systems,'' \emph{Proceedings of the IEEE}, vol. 103, no.~8, pp. 1379--1397,
  2015.

\bibitem{behdad_artificial_2015}
R.~Behdad, S.~Binczak, A.~S. Dmitrichev, V.~I. Nekorkin, and J.-M. Bilbault,
  ``Artificial {Electrical} {Morris}-{Lecar} {Neuron},'' \emph{IEEE
  Transactions on Neural Networks and Learning Systems}, vol.~26, no.~9, pp.
  1875--1884, Sep. 2015.

\bibitem{chua_memristor-missing_1971}
L.~Chua, ``Memristor-{The} {Missing} {Circuit} {Element},'' \emph{IEEE
  Transactions on Circuit Theory}, vol.~18, no.~5, pp. 507--519, 1971.

\bibitem{ochs_wave_2017}
K.~Ochs, E.~Hernandez-Guevara, and E.~Solan, ``\BIBforeignlanguage{en}{Wave
  {Digital} {Emulation} of {Spike}-{Timing} {Dependent} {Plasticity}}.''\hskip
  1em plus 0.5em minus 0.4em\relax Boston: IEEE MWSCAS, 2017.

\bibitem{blowers_energy-efficient_2014}
X.~Wu, V.~Saxena, and K.~A. Campbell, ``Energy-efficient {STDP}-based learning
  circuits with memristor synapses,'' M.~Blowers and J.~Williams, Eds., May
  2014, p. 911906.

\bibitem{ebong_cmos_2012}
I.~E. Ebong and P.~Mazumder, ``{CMOS} and {Memristor}-{Based} {Neural}
  {Network} {Design} for {Position} {Detection},'' \emph{Proceedings of the
  IEEE}, vol. 100, no.~6, pp. 2050--2060, Jun. 2012.

\bibitem{pershin_neuromorphic_2011}
Y.~V. Pershin and M.~Di~Ventra, ``Neuromorphic, {Digital} and {Quantum}
  {Computation} with {Memory} {Circuit} {Elements},'' \emph{Proceedings of the
  IEEE}, vol. 100, no.~6, pp. 2071--2080, Sep. 2011.

\bibitem{kim_memristor_2012}
H.~Kim, M.~P. Sah, C.~Yang, T.~Roska, and L.~O. Chua, ``Memristor {Bridge}
  {Synapses},'' \emph{Proceedings of the IEEE}, vol. 100, no.~6, pp.
  2061--2070, Jun. 2012.

\bibitem{ochs_wave_2017_02}
K.~Ochs, E.~Hernandez-Guevara, and E.~Solan, ``\BIBforeignlanguage{en}{Wave
  {Digital} {Information} {Anticipator}}.''\hskip 1em plus 0.5em minus
  0.4em\relax Boston: IEEE MWSCAS, 2017.

\bibitem{ochs_anticipation_2017}
K.~Ochs, M.~Ziegler, E.~Hernandez-Guevara, E.~Solan, M.~Ignatov, M.~Hansen,
  M.~S. Gill, and H.~Kohlstedt, ``\BIBforeignlanguage{en}{Anticipation of
  digital patterns},'' \emph{\BIBforeignlanguage{en}{arXiv:1704.07102 [cs]}},
  Apr. 2017, arXiv: 1704.07102.

\bibitem{strukov_missing_2008}
D.~B. Strukov, G.~S. Snider, D.~R. Stewart, and R.~S. Williams, ``The missing
  memristor found,'' \emph{Nature}, vol. 453, no. 7191, pp. 80--83, May 2008.

\bibitem{chua_memristive_1976}
L.~O. Chua and S.~M. Kang, ``Memristive {Devices} and {Systems},''
  \emph{Proceedings of the IEEE}, vol.~64, no.~2, pp. 209--223, 1976.

\bibitem{hansen_double_2015}
M.~Hansen, M.~Ziegler, L.~Kolberg, R.~Soni, S.~Dirkmann, T.~Mussenbrock, and
  H.~Kohlstedt, ``A double barrier memristive device,'' \emph{Scientific
  Reports}, vol.~5, p. 13753, Sep. 2015.

\bibitem{ochs_sensitivity_2016}
K.~Ochs and E.~Solan, ``\BIBforeignlanguage{en}{Sensitivity analysis of
  memristors based on emulation techniques},'' in
  \emph{\BIBforeignlanguage{en}{2016 {IEEE} 59th {International} {Midwest}
  {Symposium} on {Circuits} and {Systems} ({MWSCAS})}}.\hskip 1em plus 0.5em
  minus 0.4em\relax Abu Dhabi, UAE: IEEE MWSCAS, 2016, pp. 1--4.

\bibitem{biolek_reliable_2013}
D.~Biolek, M.~Di~Ventra, and Y.~V. Pershin, ``Reliable {SPICE} {Simulations} of
  {Memristors}, {Memcapacitors} and {Meminductors},'' \emph{Radioengineering},
  vol.~22, no. 945, Jul. 2013.

\bibitem{sah_generic_2015}
M.~P. Sah, C.~Yang, H.~Kim, B.~Muthuswamy, J.~Jevtic, and L.~Chua, ``A
  {Generic} {Model} of {Memristors} {With} {Parasitic} {Components},''
  \emph{IEEE Transactions on Circuits and Systems I: Regular Papers}, vol.~62,
  no.~3, pp. 891--898, Mar. 2015.

\bibitem{solan_enhanced_2017}
E.~Solan, S.~Dirkmann, M.~Hansen, D.~Schroeder, H.~Kohlstedt, M.~Ziegler,
  {Thomas Mussenbrock}, and K.~Ochs, ``\BIBforeignlanguage{en}{An enhanced
  lumped element electrical model of a double barrier memristive device},''
  \emph{\BIBforeignlanguage{en}{Journal of Physics D: Applied Physics}},
  vol.~50, no.~19, p. 195102, 2017.

\bibitem{nguyen_total_2013}
T.~Nguyen, ``Total {Number} of {Synapses} in the {Adult} {Human} {Neocortex},''
  \emph{Undergraduate Journal of Mathematical Modeling: One + Two}, vol.~3,
  no.~1, May 2013.

\bibitem{hyongsuk_kim_memristor_2012}
{Hyongsuk Kim}, M.~P. Sah, {Changju Yang}, {Seongik Cho}, and L.~O. Chua,
  ``Memristor {Emulator} for {Memristor} {Circuit} {Applications},'' \emph{IEEE
  Transactions on Circuits and Systems I: Regular Papers}, vol.~59, no.~10, pp.
  2422--2431, Oct. 2012.

\bibitem{sanchez-lopez_860khz_2017}
C.~Sánchez-López and L.~Aguila-Cuapio, ``\BIBforeignlanguage{en}{A 860khz
  grounded memristor emulator circuit},'' \emph{\BIBforeignlanguage{en}{AEU -
  International Journal of Electronics and Communications}}, vol.~73, pp.
  23--33, Mar. 2017.

\bibitem{biolek_passive_2015}
D.~Biolek, V.~Biolkova, Z.~Kolka, and Z.~Biolek, ``Passive fully floating
  emulator of memristive device for laboratory experiments,'' \emph{Advances in
  Electrical and Computer Engineering}, 2015.

\bibitem{triverio_stability_2007}
P.~Triverio, S.~Grivet-Talocia, M.~Nakhla, F.~Canavero, and R.~Achar,
  ``Stability, {Causality}, and {Passivity} in {Electrical} {Interconnect}
  {Models},'' \emph{IEEE Transactions on Advanced Packaging}, vol.~30, no.~4,
  pp. 795--808, Nov. 2007.

\bibitem{fettweis_wave_1986}
A.~Fettweis, ``Wave digital filters: {Theory} and practice,'' \emph{Proceedings
  of the IEEE}, vol.~74, no.~2, pp. 270--327, Feb. 1986.

\bibitem{ochs_passive_2001}
K.~Ochs, ``Passive integration methods: {Fundamental} theory,'' \emph{Intern.
  Journal of Electronics and Communications (AEÜ)}, vol.~55, no.~3, pp.
  153--163, May 2001.

\bibitem{fettweis_pseudo-passivity_1972}
A.~Fettweis, ``Pseudo-passivity, sensitivity, and stability of wave digital
  filters,'' \emph{IEEE Transactions on Circuit Theory}, vol.~19, no.~6, pp.
  668--673, 1972.

\bibitem{schwerdtfeger_multidimensional_2014}
\BIBentryALTinterwordspacing
T.~Schwerdtfeger and A.~Kummert, ``A multidimensional signal processing
  approach to wave digital filters with topology-related delay-free loops,'' in
  \emph{Acoustics, {Speech} and {Signal} {Processing} ({ICASSP}), 2014 {IEEE}
  {International} {Conference} on}.\hskip 1em plus 0.5em minus 0.4em\relax
  IEEE, 2014, pp. 389--393. [Online]. Available:
  \url{http://ieeexplore.ieee.org/abstract/document/6853624/}
\BIBentrySTDinterwordspacing

\bibitem{ochs_wave_2016_02}
K.~Ochs and E.~Solan, ``\BIBforeignlanguage{en}{Wave digital emulation of
  charge-or flux-controlled memristors},'' in
  \emph{\BIBforeignlanguage{en}{2016 {IEEE} 59th {International} {Midwest}
  {Symposium} on {Circuits} and {Systems} ({MWSCAS})}}.\hskip 1em plus 0.5em
  minus 0.4em\relax Abu Dhabi, UAE: IEEE MWSCAS, 2016, pp. 1--4.

\bibitem{chua_resistance_2011}
L.~Chua, ``\BIBforeignlanguage{en}{Resistance switching memories are
  memristors},'' \emph{\BIBforeignlanguage{en}{Applied Physics A}}, vol. 102,
  no.~4, pp. 765--783, Mar. 2011.

\bibitem{ho_dynamical_2011}
Y.~Ho, G.~M. Huang, and P.~Li, ``Dynamical {Properties} and {Design} {Analysis}
  for {Nonvolatile} {Memristor} {Memories},'' \emph{IEEE Transactions on
  Circuits and Systems I: Regular Papers}, vol.~58, no.~4, pp. 724--736, Apr.
  2011.

\bibitem{dirkmann_kinetic_2015}
S.~Dirkmann, M.~Ziegler, M.~Hansen, H.~Kohlstedt, J.~Trieschmann, and
  T.~Mussenbrock, ``\BIBforeignlanguage{en}{Kinetic simulation of filament
  growth dynamics in memristive electrochemical metallization devices},''
  \emph{\BIBforeignlanguage{en}{Journal of Applied Physics}}, vol. 118, no.~21,
  p. 214501, Dec. 2015.

\bibitem{waser_redox-based_2009}
R.~Waser, R.~Dittmann, G.~Staikov, and K.~Szot,
  ``\BIBforeignlanguage{en}{Redox-{Based} {Resistive} {Switching} {Memories} -
  {Nanoionic} {Mechanisms}, {Prospects}, and {Challenges}},''
  \emph{\BIBforeignlanguage{en}{Advanced Materials}}, vol.~21, no. 25-26, pp.
  2632--2663, Jul. 2009.

\bibitem{thomas_memristor-based_2013}
A.~Thomas, ``Memristor-based neural networks,'' \emph{Journal of Physics D:
  Applied Physics}, vol.~46, no.~9, p. 093001, Mar. 2013.

\bibitem{wang_sericin_2013}
H.~Wang, F.~Meng, Y.~Cai, L.~Zheng, Y.~Li, Y.~Liu, Y.~Jiang, X.~Wang, and
  X.~Chen, ``\BIBforeignlanguage{en}{Sericin for {Resistance} {Switching}
  {Device} with {Multilevel} {Nonvolatile} {Memory}},''
  \emph{\BIBforeignlanguage{en}{Advanced Materials}}, vol.~25, no.~38, pp.
  5498--5503, Oct. 2013.

\bibitem{zhou_multilevel_2016}
W.~Zhou, Y.~Xiong, Z.~Zhang, D.~Wang, W.~Tan, Q.~Cao, Z.~Qian, and Y.~Du,
  ``\BIBforeignlanguage{en}{Multilevel {Resistance} {Switching} {Memory} in
  {La} $_{\textrm{2/3}}$ {Ba} $_{\textrm{1/3}}$ {MnO} $_{\textrm{3}}$
  /0.7pb({Mg} $_{\textrm{1/3}}$ {Nb} $_{\textrm{2/3}}$ ){O} $_{\textrm{3}}$
  -0.3pbtio $_{\textrm{3}}$ (011) {Heterostructure} by {Combined}
  {Straintronics}-{Spintronics}},'' \emph{\BIBforeignlanguage{en}{ACS Applied
  Materials \& Interfaces}}, vol.~8, no.~8, pp. 5424--5431, Mar. 2016.

\bibitem{dirkmann_role_2016}
S.~Dirkmann, M.~Hansen, M.~Ziegler, H.~Kohlstedt, and T.~Mussenbrock, ``The
  role of ion transport phenomena in memristive double barrier devices,''
  \emph{Scientific Reports}, vol.~6, no. 35686, pp. 1--12, Oct. 2016.

\bibitem{simmons_generalized_1963}
\BIBentryALTinterwordspacing
J.~G. Simmons, ``\BIBforeignlanguage{en}{Generalized {Formula} for the
  {Electric} {Tunnel} {Effect} between {Similar} {Electrodes} {Separated} by a
  {Thin} {Insulating} {Film}},'' \emph{\BIBforeignlanguage{en}{Journal of
  Applied Physics}}, vol.~34, no.~6, pp. 1793--1803, Jan. 1963. [Online].
  Available:
  \url{http://scitation.aip.org/content/aip/journal/jap/34/6/10.1063/1.1702682}
\BIBentrySTDinterwordspacing

\bibitem{ochs_wave_2016}
K.~Ochs, E.~Solan, S.~Dirkmann, and T.~Mussenbrock,
  ``\BIBforeignlanguage{en}{Wave digital emulation of a double barrier
  memristive device},'' in \emph{\BIBforeignlanguage{en}{2016 {IEEE} 59th
  {International} {Midwest} {Symposium} on {Circuits} and {Systems}
  ({MWSCAS})}}.\hskip 1em plus 0.5em minus 0.4em\relax Abu Dhabi, UAE: IEEE
  MWSCAS, 2016, pp. 1--4.

\bibitem{solan_parameter_2017}
E.~Solan and K.~Ochs, ``\BIBforeignlanguage{en}{Parameter {Identification} of a
  {Double} {Barrier} {Memristive} {Device}}.''\hskip 1em plus 0.5em minus
  0.4em\relax Boston: IEEE MWSCAS, 2017.

\bibitem{ochs_consistent_2017}
K.~Ochs and E.~Solan, ``A {Consistent} {Modeling} of {Passive} {Memcapacitive}
  {Systems}.''\hskip 1em plus 0.5em minus 0.4em\relax Boston: IEEE MWSCAS,
  2017.

\end{thebibliography}

%








\end{document}